\documentclass[a4paper,fleqn,usenatbib]{mnras}

\usepackage{newtxtext,newtxmath}

\usepackage[T1]{fontenc}
\usepackage{ae,aecompl}

\usepackage{graphicx}	
\usepackage{amsmath}	

\newcommand{\vsini}{$v_{\rm e} \sin i$}
\newcommand{\kms}{km\,s$^{-1}$}

\newcommand{\bz}{$\langle B_{\rm z}\rangle$}
\newcommand{\her}{45~Her}

\newcommand{\figps}[3]{\resizebox{#1}{!}{\rotatebox{#2}{\includegraphics{#3}}}}

\title[Magnetic field and surface structure of 45~Her]{Surface structure of 45 Hercules: An otherwise unremarkable Ap star with a surprisingly weak magnetic field}

\author[O. Kochukhov et al.]
{O.\ Kochukhov$^1$\thanks{E-mail: oleg.kochukhov@physics.uu.se},
H.\ G\"ursoytrak Mutlay$^2$,
A.\ M.\ Amarsi$^1$,
P.\ Petit$^3$,
I.\ Mutlay$^4$,
B.\ G\"urol$^2$
\\
$^1$Department of Physics and Astronomy, Uppsala University, Box 516, Uppsala 75120, Sweden \\
$^2$Department of Astronomy and Space Sciences, Faculty of Science, Ankara University, 06100, Ankara, T\"urkiye \\
$^3$Institut de Recherche en Astrophysique et Plan\'etologie, Universit\'e de Toulouse, CNRS, CNES, 14 avenue \'Edouard Belin, 31400 Toulouse, France\\
$^4$Department of Chemical Engineering, Faculty of Engineering, Ankara University, 06100, Ankara, T\"urkiye
}

\date{Accepted 2023 March 6. Received 2023 March 6; in original form 2023 February 2}

\pubyear{2023}

\begin{document}
\label{firstpage}
\pagerange{\pageref{firstpage}--\pageref{lastpage}}
\maketitle

\begin{abstract}
The origin of magnetic fields and their role in chemical spot formation on magnetic Ap stars is currently not understood. Here we contribute to solving this problem with a detailed observational characterisation of the surface structure of 45~Her, a weak-field Ap star. We find this object to be a long-period, single-lined spectroscopic binary and determine the binary orbit as well as fundamental and atmospheric parameters of the primary. We study magnetic field topology and chemical spot distribution of 45~Her with the help of the Zeeman Doppler imaging technique. Magnetic mapping reveals the stellar surface field to have a distorted dipolar topology with a surface-averaged field strength of 77~G and a dipolar component strength of 119~G -- confirming it as one of the weakest well-characterised Ap-star fields known. Despite its feeble magnetic field, 45~Her shows surface chemical inhomogeneities with abundance contrasts of up to 6 dex. Of the four chemical elements studied, O concentrates at the magnetic equator whereas Ti, Cr and Fe avoid this region. Apart from this trend, the positions of Fe-peak element spots show no apparent correlation with the magnetic field geometry. No signs of surface differential rotation or temporal evolution of chemical spots on the time scale of several years were detected. Our findings demonstrate that chemical spot formation does not require strong magnetic fields to proceed and that both the stellar structure and the global field itself remain stable for sub-100~G field strengths contrary to theoretical predictions.
\end{abstract}




\section{Introduction}
\label{sec:intro}

About 10 per cent of main sequence A and B-type stars show surface magnetic fields with strengths ranging from a few hundred G up to $\sim$\,30~kG \citep[e.g.][]{donati:2009,sikora:2019}. These magnetic fields exhibit a relatively simple surface structure, approximately dipolar in the majority of stars, that remains static on the observable time scales. Neither the origin nor the incidence of these so-called `fossil' magnetic fields is currently understood \citep[e.g.][]{neiner:2015}. Generally, these fields are believed to be hydrodynamically stable \citep{braithwaite:2004} remnants of one or several magnetic field generation processes operating at an earlier stellar evolutionary stage \citep{mestel:2003,moss:2004,schneider:2019}.

The presence of kG-strength magnetic fields in a tenuous atmosphere of an upper main sequence star has a drastic impact on the surface structure and atmospheric physics. Most importantly, various chemical elements separate under the influence of competing processes of radiative levitation and gravitational settling, forming non-uniform horizontal and vertical distributions. This, in turn, gives rise to the observed extreme spectral peculiarities of Ap/Bp (also known as the upper main sequence chemically peculiar or CP, see \citealt{preston:1974}) stars and their spot-induced rotational variability. This combination of unique characteristics makes magnetic Ap stars prime targets for investigations of non-trivial interactions between magnetic fields and stellar plasma with both observational (e.g. \citealt{kochukhov:2017}, \citealt*{kochukhov:2019} and references therein) and theoretical \citep*[e.g.][]{michaud:1981,babel:1991,alecian:2015} approaches. These efforts notwithstanding, no successful theoretical model explaining the observed diversity of chemical spot topologies has emerged and the very foundation of the magnetically-driven element segregation hypothesis has been challenged by the discovery of time-dependent chemical spot formation in HgMn stars \citep[e.g.][]{kochukhov:2007b,kochukhov:2011b,kochukhov:2021b}, which lack organised magnetic fields typical of Ap stars. In this context, it is of particular interest to assess the relationship between magnetic field and chemical spots for well-established magnetic Ap stars at the lower boundary of the range of surface field strength probed by these objects. Are these weak-field Ap stars similar or different compared to stars with stronger fields? Is there any evidence of a systematic reduction of spot contrasts or emergence of temporal variation of spot topologies with the decrease of magnetic field strength? Are the observed field strengths compatible with the theoretically predicted thresholds of fossil field stability?

The target of the present study -- the magnetic Ap star \her\ (l Her, V776 Her, HR 6234, HD 151525) -- is uniquely suited to address some of these questions. This bright ($V=5.24$ mag) star is given a B9p CrEu spectral classification in the literature \citep{renson:2009} and is known to be spectral \citep{deutsch:1947} and photometric \citep{burke:1981} variable with a period of $\approx$\,4.12~d \citep[e.g.][]{paunzen:2021b}. Quantitative modelling of this variability was attempted only once by \citet{hatzes:1991a}, who derived a surface equivalent width map from a single \ion{Cr}{ii} spectral line. 
What sets \her\ apart from more typical magnetic Ap stars is its illusive magnetic field. Although this star was suspected to be magnetic already by \citet{babcock:1958a}, the first definitive magnetic field measurement was reported by \citet{auriere:2007}. Using relatively low quality spectropolarimetric data, these authors found a dipolar field strength of $0.54\substack{+1.38\\-0.34}$ kG, suggesting that \her\ might host one of the weakest surface magnetic fields among Ap stars. No information about higher-order magnetic field components or the spatial relationship between the stellar magnetic field topology and chemical spot distributions is available in the literature.

In this study we present the first detailed magnetic and chemical abundance study of the surface structure of \her\ based on a large, high quality spectropolarimetric data set as well as two independent spectroscopic time series collected at different observatories over the time span of four years. This analysis allowed us to gain significant new insights into the surface physics of weak-field Ap stars. The rest of this paper is organised as follows. Sect.~\ref{sec:obs} describes observational data employed in our study. Sect.~\ref{sec:res} presents investigation of the spectroscopic binary orbit (Sect.~\ref{sec:binary}), determination of stellar parameters and mean abundances (Sect.~\ref{sec:params}), derivation of the mean intensity and polarisation spectral line profiles (Sect.~\ref{sec:lsd}), and their application for surface structure mapping (Sect.~\ref{sec:zdi}). The paper is concluded with a summary and discussion of our main findings in Sect.~\ref{sec:disc}.

\section{Observational data}
\label{sec:obs}

\subsection{Narval spectropolarimetry}

High-resolution spectropolarimetric observations of \her\ analysed in this study were obtained with the Narval instrument installed at the 2-m T\'elescope Bernard Lyot of the Pic du Midi Observatory. Narval is a cross-dispersed \'echelle, dual-beam spectropolarimeter covering the wavelength range 360--1000~nm with a resolution of $R\approx65000$. Our target was observed 46 times, with two observations acquired in March 2007 followed by an intense monitoring campaign between April and October 2018, which yielded the remaining 44 observations. Circular polarisation (Stokes $V$) spectra were obtained from one to four times a night, using a total exposure time of 1200~s. Each such observation was split into four 300~s sub-exposures obtained with two different configurations of the polarimetric unit. As described by \citet{donati:1997} and \citet{bagnulo:2009}, this dual-beam polarimetric modulation technique allows one to remove spurious instrumental artefacts and derive diagnostic null spectra alongside the intensity and polarisation spectra. 

The main data reduction and calibration steps were accomplished with an updated version of the {\sc esprit} code \citep{donati:1997} executed automatically at the telescope. The reduction products produced by this pipeline are available in the PolarBase\footnote{\url{http://polarbase.irap.omp.eu}} archive \citep{petit:2014}. We considered the non-normalised version of the spectra and performed continuum rectification with the methods described by \citet{rosen:2018}. The journal of Narval observations of \her\ is provided in Table~\ref{tab:obs}, which gives the UT date of each observation, the heliocentric Julian date of mid-exposure, and the signal-to-noise ratio ($S/N$) in the first three columns. This $S/N$ refers to one pixel, approximately spanning a 1.8~\kms\ velocity bin, of the extracted Stokes $V$ spectrum at $\lambda$\,=\,550~nm. It was calculated from the standard deviation of the null spectrum and ranges from around 100 to nearly 1000 with a median value of 529.

\subsection{CES spectroscopy}

Spectroscopic observations of \her\ were obtained at the T\"UBITAK National Observatory (TUG), Antalya with RTT150 (1.5-meter Russian-Turkish) telescope \citep{aslan:2001} and Coude Echelle Spectrometer (CES). Observations were carried out on 13 different nights in 2014, 2015, and 2018 with exposure times ranging from 20 to 70 min. In total, 48 spectra of \her\ with $S/N$ in the 100--200 range were obtained. CES spectra have a resolution of 40000 
and cover a wavelength interval from 369~nm to 1028~nm. Because of telluric line contamination, low $S/N$, and continuum normalisation problems, only the 400--700~nm region was used in this work. 

In each observing night, we obtained ten biases, five dome flats, and three ThAr (thorium-argon) arc lamp spectra for calibrating the raw spectra of \her. All spectral reduction, wavelength calibration, and spectral normalisation step were performed with IRAF \citep[Image Reduction and Analysis Facility,][]{tody:1986} procedures. The $S/N$ was determined using IRAF, by considering the scatter of spectral points in the line-free regions around $\lambda$\,=\,550~nm. The average size of CES spectral pixel is 2.4~\kms.

The journal of CES observations of \her\ is given in Table~\ref{tab:obsrtt}, which reports the UT and heliocentric Julian dates for each observation, exposure time, achieved $S/N$, the rotational and orbital phases.

\subsection{Spectrophotometry}

Different spectrophotometric observations of \her\ were collected from the literature for the purpose of modelling the stellar spectral energy distribution (SED) from ultra-violet to near-infrared. The UV part of the stellar SED is covered by the TD-1 satellite measurements \citep{thompson:1978}. For the 336--1020~nm wavelength region we used the newly released Gaia DR3 externally calibrated spectrophotometry \citep{montegriffo:2022}. We also considered Geneva photometry \citep{rufener:1988} converted to absolute fluxes at 7 wavelength bands according to the prescription by \citet{rufener:1988a}. In the near-infrared, 2MASS photometry \citep{cutri:2003} as well as JHKLM observations by \citet{groote:1983} are available. The magnitudes provided by these two catalogues were converted to absolute flux units using the calibrations by \citet{cohen:2003} and \citet*{van-der-bliek:1996}, respectively. The resulting composite observed SED of \her\ is illustrated in Fig.~\ref{fig:sed}.

\section{Analysis and results}
\label{sec:res}

\subsection{Spectroscopic binary orbit}
\label{sec:binary}

An initial look at the Narval spectra of \her\ revealed the presence of $\approx$\,10~\kms\ peak-to-peak radial velocity changes on a time scale significantly longer than the stellar rotational period. Although some mentions of possible spectroscopic binarity of \her\ can be found in the literature \citep{mason:2001}, studies focused on multiplicity analysis were unable to confirm radial velocity variations and derive orbital parameters \citep{carrier:2002,chini:2012}. Similar to many other Ap stars, the low-amplitude radial velocity variation of metal lines in \her\ spectrum is dominated by the rotational variability produced by chemical abundance inhomogeneities. In this case, chemical spots are responsible for the radial velocity modulation between $\pm$3 and $\pm$6~\kms, depending on chemical element. This prevents a straightforward measurement of the orbital radial velocity variation. However, as demonstrated by previous studies (e.g. \citealt{bischoff:2017}; \citealt*{kochukhov:2022}), radial velocity measurements using hydrogen lines are nearly insensitive to heavy element spots and can be successfully employed to investigate the orbital motion of Ap stars.

In this study we obtained radial velocity measurements from the centre-of-gravity positions of the cores of hydrogen H$\alpha$ to H$\delta$ lines. The resulting average radial velocity is reported for Narval spectra in the last column of Table~\ref{tab:obs}. 
For the lower quality CES spectra we found that better results are obtained using the core of H$\alpha$ and averaging measurements for each night when multiple spectra were taken. The resulting 13 average radial velocities are given in Table~\ref{tab:rttrv}. 

\begin{figure}
\centering
\figps{\hsize}{0}{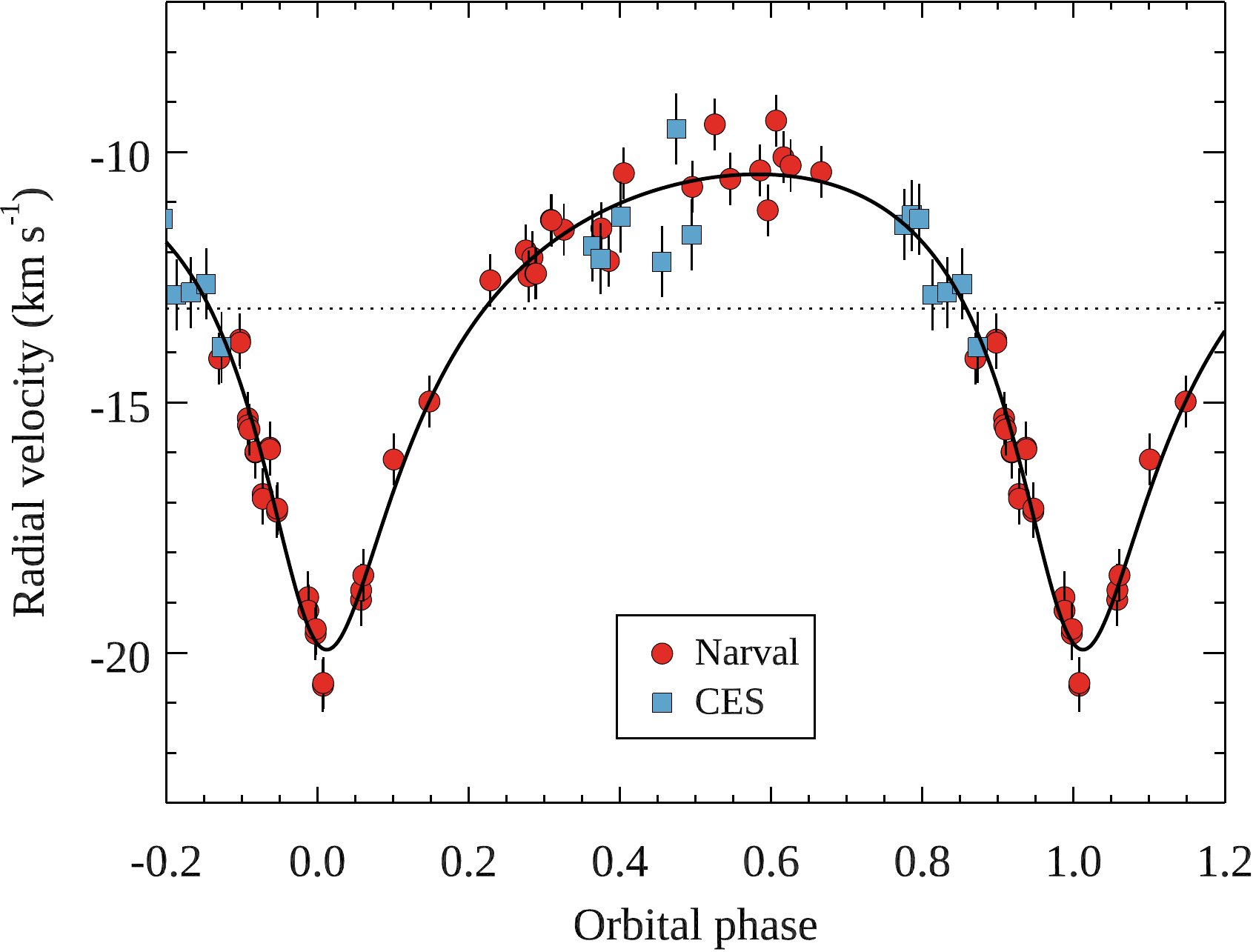}
\caption{Orbital radial velocity variation of \her. Symbols show the radial velocity measurements derived from the Narval (circles) and CES (squares) observations. The solid line corresponds to the best-fitting orbital solution. The dotted line shows systemic velocity.}
\label{fig:orbit}
\end{figure}

We fitted a set of SB1 orbital elements ($P_{\rm orb}$, HJD$_0$, $K_1$, $\gamma$, $e$, $\omega$) to the combined Narval and CES data set using a non-linear least-squares optimisation algorithm \citep{markwardt:2009}. Figure~\ref{fig:orbit} compares the observed radial velocities with the best-fitting orbital solution. The corresponding orbital parameters are given in Table~\ref{tab:orbit} and the orbital phases of Narval observations are reported in the fifth column of Table~\ref{tab:obs}. The scatter of the average hydrogen-line radial velocity measurements around the orbital solution amounts to 0.52~\kms\ for individual Narval observations and 0.71~\kms\ for nightly averaged CES spectra, which we take as an estimate of the measurement uncertainties for the two data sets.

This analysis establishes \her\ as the primary in a spectroscopic binary system with $P_{\rm orb}$ of about 100~d, significant eccentricity and a semi-amplitude of only 4.8~\kms. The radial velocity shifts due to this orbital motion were removed before further spectroscopic and magnetic analysis. The mass function $f$ corresponding to the derived orbital solution implies a minimum mass of the companion $M_2=0.193\pm0.013$\,$M_\odot$ if $M_1=2.84\pm0.13$\,$M_\odot$ (see Sect.~\ref{sec:params}) is adopted for the primary. If one further assumes the orbital and rotational axes to be aligned ($i_{\rm orb}=i_{\rm rot}=42\degr$, see Sect.~\ref{sec:params}), the mass of the secondary is $0.295\pm0.018$\,$M_\odot$. Thus, the secondary in the \her\ system is most likely a late-type low-mass star that makes a negligible contribution to the composite high-resolution optical spectrum or a wide-band SED.

\begin{table}
\caption{Spectroscopic orbital parameters and mass function for \her. \label{tab:orbit}}
\begin{tabular}{ll}
\hline
Parameter & Value \\
\hline
%
$P_{\rm orb}$  (d) &   $99.51\pm0.10$ \\
HJD$_0$ (d) & $2456702.6\pm2.1$ \\
$K_1$ (\kms) & $4.75\pm0.12$ \\
$\gamma$ (\kms) & $-13.13\pm0.09$ \\
$e$  &     $0.445\pm0.020$ \\
$\omega$ (\degr) &  $167.1\pm4.2$ \\
$f$ ($M_\odot$) & $(7.92\pm0.67)\times10^{-4}$ \\
\hline
\end{tabular}
\end{table}

\subsection{Stellar parameters and mean abundances}
\label{sec:params}

\begin{figure*}
\figps{\hsize}{90}{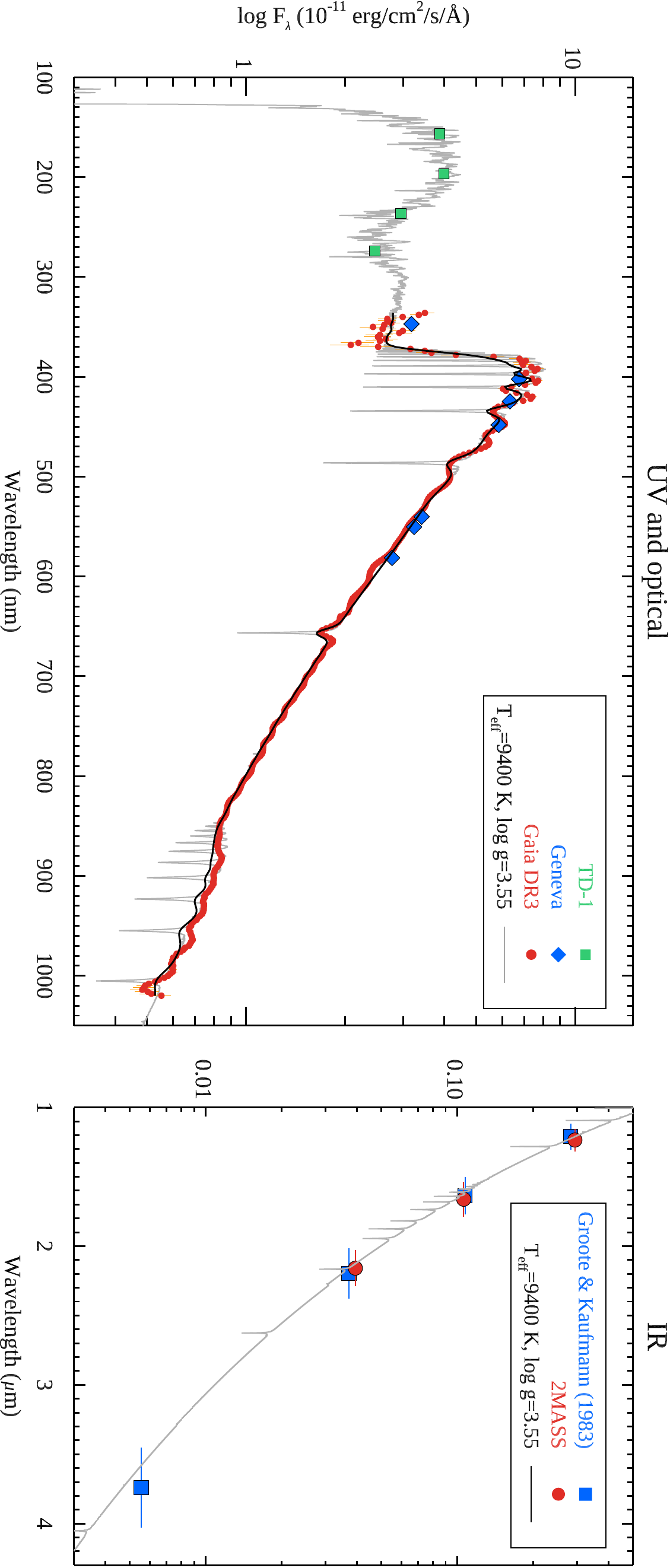}
\caption{Comparison of the observed and computed spectrophotometry of \her\ in the UV and optical (left panel) and near-infrared (right panel). Different sources of the observed spectrophotometric data are indicted in the legends. The grey solid line shows the best-fitting theoretical spectral energy distribution for $T_{\rm eff}=9400$~K, $\log g=3.55$, $R=4.3 R_\odot$, $E(B-V)=0.02$ convolved with a constant Gaussian profile. The black solid line shows the same theoretical model spectrum convolved with a wavelength-dependent broadening function \citep[see][]{montegriffo:2022} appropriate for the Gaia spectrophotometry.}
\label{fig:sed}
\end{figure*}

\begin{figure}
\figps{\hsize}{0}{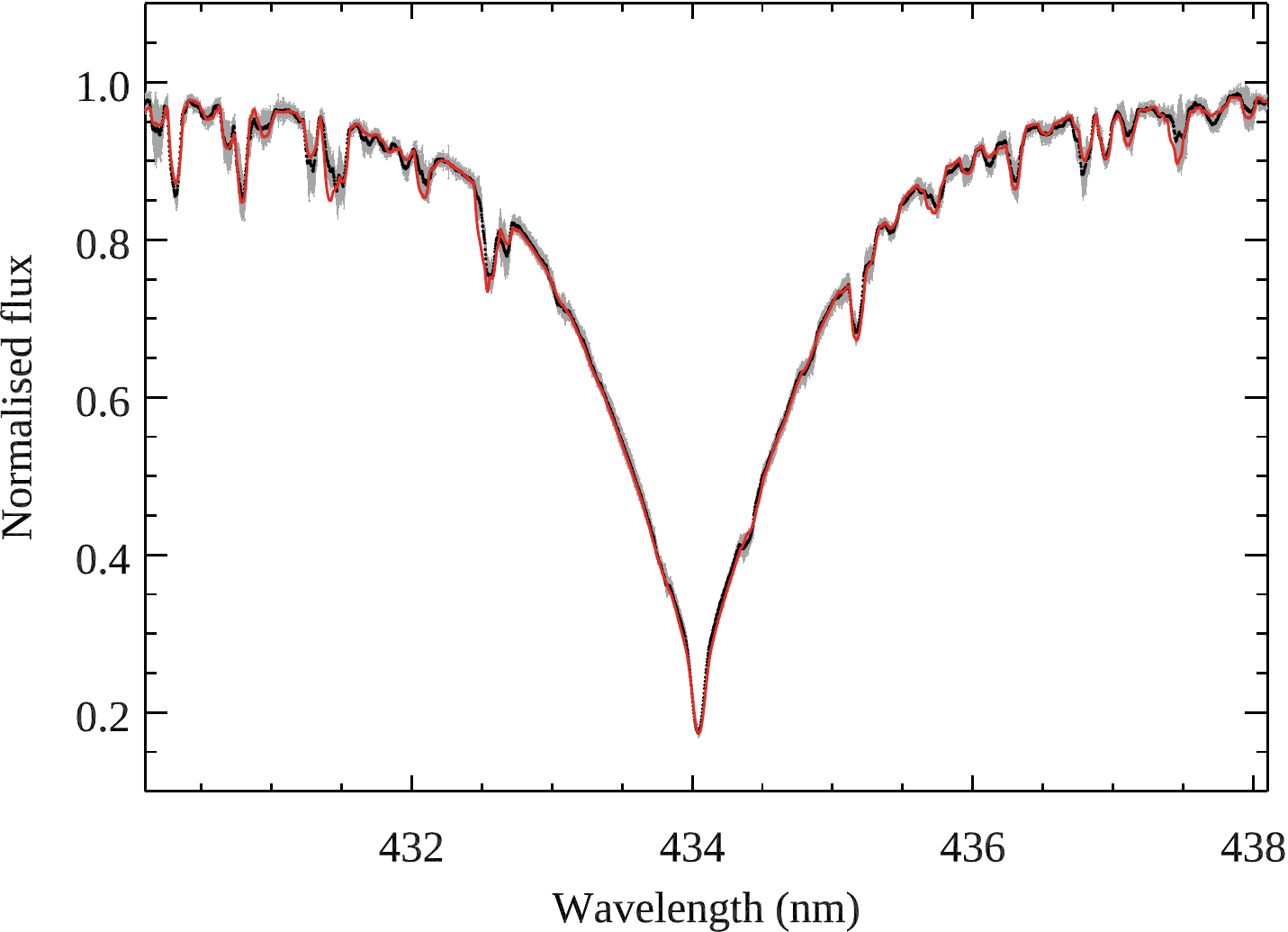}
\caption{Comparison of the average observed (black symbols) and computed (red solid line) hydrogen H$\gamma$ profile of \her. The grey background curve corresponds to the average spectrum $\pm$ one standard deviation.}
\label{fig:hline}
\end{figure}

\begin{figure}
\centering
\figps{\hsize}{0}{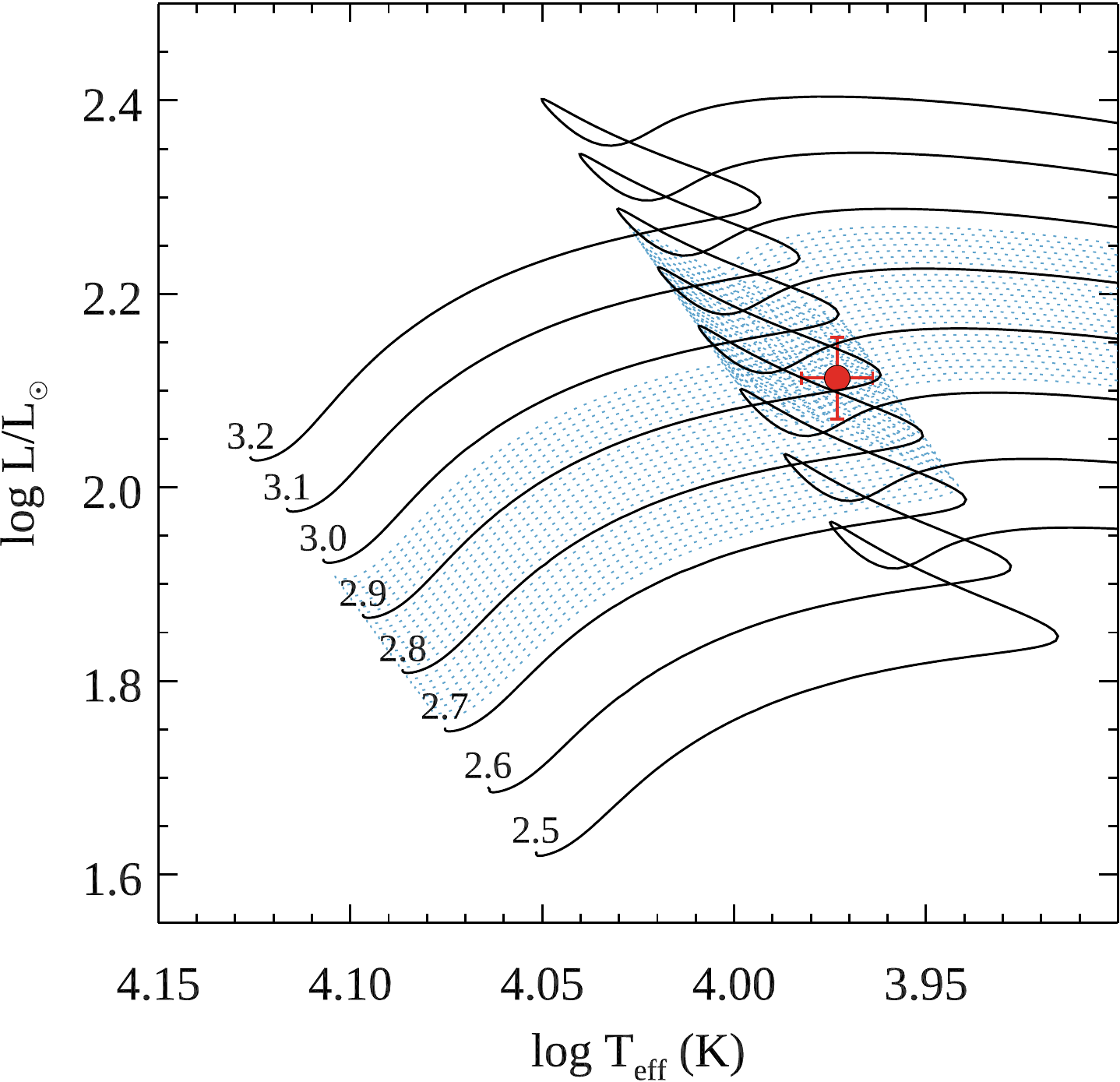}
\caption{Temperature and luminosity of \her\ compared to MIST theoretical evolutionary tracks for the initial stellar masses 2.5--3.2 $M_\odot$. Blue dotted lines show the set of evolutionary tracks compatible with $T_{\rm eff}$ and $L$ of \her.}
\label{fig:hrd}
\end{figure}

Multiple photometric determinations of the atmospheric parameters of \her\ can be found in the literature \citep[e.g.][]{kochukhov:2006,auriere:2007,netopil:2008,netopil:2017,glagolevskij:2019}. These analyses reported $T_{\rm eff}$\,=\,9200--9400~K and $\log g$\,=\,3.5--3.8. Based on these results, we carried out an analysis of the mean composition of \her\ using phase-averaged spectra. We used {\sc LLmodels} \citep{shulyak:2004} model atmosphere with $T_{\rm eff}$\,=\,9250~K, $\log g$\,=\,3.5, $[\mathrm{M/H}]=+0.5$. Abundances were determined using VALD \citep{ryabchikova:2015} line lists and {\sc Synth3} \citep{kochukhov:2007d} spectrum synthesis code run from within {\sc BinMag} \citep{kochukhov:2018} fitting and visualisation interface. Typically, one to three element abundances together with the projected rotational velocity were determined by fitting theoretical calculations to small segments in the average Narval spectrum of \her. 
In this analysis, the synthetic spectra were computed under the local thermodynamic equilibrium (LTE) assumption, and therefore lines known to show strong departures from LTE, e.g. the \ion{O}{i} near-infrared triplet, were avoided. The resulting abundances are summarised in Table~\ref{tab:abn}. The average projected rotational velocity emerging from this analysis is \vsini\,=\,$36.8\pm1.4$~\kms.  

Comparing \her\ abundances to solar values, we find that this object shows a typical Ap-star abundance pattern \citep[e.g.][]{ryabchikova:2005b} with a moderate underabundance of light elements and significant overabundance of iron-peak elements, particularly Cr. A somewhat less common feature relative to Ap stars with similar $T_{\rm eff}$ is the lack of any Si overabundance and a noticeable depletion of Ti. Our analysis reveals no conclusive evidence of large overabundances of rare-earth or other heavy elements -- a feature previously seen in other old Ap stars (\citealt*{ryabchikova:2005a}; \citealt{kochukhov:2006b}).

Based on this abundance analysis, we computed a custom grid of {\sc LLmodels} atmospheres in order to refine $T_{\rm eff}$ and $\log g$. The effective temperature was determined with a global fit to the observed SED (Fig.~\ref{fig:sed}). The stellar angular diameter would normally be another parameter derived in this fit. Considering availability of a relatively accurate Gaia DR3 parallax, $7.72 \pm0.11$~mas \citep{gaia-collaboration:2021}, one can replace the angular diameter with the stellar radius in SED modelling. At the same time, results of SED fitting are also influenced by interstellar extinction and reddening. Three-dimensional interstellar medium maps compiled by \citet{lallement:2019} predict $E(B-V)=0.022\pm0.019$, suggesting a small and rather uncertain reddening. Owing to availability of the UV spectrophotometry, the colour excess can be independently determined from the SED fit simultaneously with $T_{\rm eff}$ and $R$. We found that the set of parameters yielding the best description of the observed SED is $T_{\rm eff}$\,=\,$9400\pm200$~K, $R$\,=\,$4.3\pm0.1$~$R_\odot$, and $E(B-V)=0.02\pm0.01$. The corresponding model SED is compared with the observed fluxes from UV to near-infrared in Fig.~\ref{fig:sed}.

The surface gravity was refined by fitting the H$\gamma$ line profile in the average Narval spectrum of \her. Theoretical Balmer line profiles were calculated with the {\sc Synth3} code for the custom model atmosphere grid described above. As demonstrated by Fig.~\ref{fig:hline}, observations are best reproduced with $\log g$\,=\,$3.55\pm0.05$.

The temperature and radius determined for \her\ correspond to luminosity $\log L/L_\odot=2.113\pm0.042$. Using this information, we examined evolutionary state and estimated the mass of \her\ by comparing its position in the Hertzsprung-Russell diagram (HRD) with the predictions of MIST\footnote{MESA Isochrones \& Stellar Tracks, where MESA stands for Modules for Experiments in Stellar Astrophysics \citep{paxton:2011}.} \citep{dotter:2016,choi:2016} solar-metallicity theoretical stellar evolutionary tracks. As evident from Fig.~\ref{fig:hrd}, \her\ is an intermediate-mass star approaching the end of its main sequence life. Since evolutionary tracks overlap in this HRD region, one cannot associate a unique mass and age to a given $T_{\rm eff}$ and $L$ pair. Instead, we considered a fine grid of model tracks, calculated with a step of 0.01 $M_\odot$, and noted all tracks passing within the one-sigma error ellipse around 45~Her's parameters (blue dotted lines in Fig.~\ref{fig:hrd}). This procedure yielded masses in the 2.71--2.97 $M_\odot$ range and ages between 351 and 469 Myr. The corresponding model tracks have zero-age main sequence radii of 1.77--1.87 $R_\odot$.

The summary of atmospheric and physical parameters of \her\ discussed in this section is given in Table~\ref{tab:params}.

\begin{table}
\caption{Mean abundances of \her\ compared to the solar composition from \citet*{asplund:2021}. Uncertain values derived from single spectral features are indicated with ':'. \label{tab:abn}}
\begin{tabular}{lll}
\hline
Element & $\log (N_{\rm el}/N_{\rm tot})$ & $\log (N_{\rm el}/N_{\rm tot})_\odot$  \\
\hline
He & $<-1.30$: & $-1.12$ \\
 C & $-4.24\pm0.37$ & $-3.57$ \\
N & $-5.13$: & $-4.20$ \\
O & $-4.12$: & $-3.34$ \\
Mg & $-4.70\pm0.21$ & $-4.48$ \\
Al & $-5.48$: & $-5.60$ \\
Si & $-4.61\pm0.42$ & $-4.52$ \\
Ca & $-6.62\pm0.46$ & $-5.73$ \\
Ti & $-7.69\pm0.31$ & $-7.06$ \\
Cr & $-4.71\pm0.20$ & $-6.41$ \\
Mn & $-5.63\pm0.16$ & $-6.61$ \\
Fe & $-4.16\pm0.15$ & $-4.57$ \\
Ba & $-9.40\pm0.11$ & $-9.76$ \\
Y & $-9.67$: & $-9.82$ \\
Sr & $-9.93$: & $-9.20$ \\
\hline
\end{tabular}
\end{table}

\subsection{Multi-line profile analysis}
\label{sec:lsd}

\subsubsection{Least-squares deconvolved profiles}

Although suspected to be magnetic already by \citet{babcock:1958a}, \her\ proved to be one of the most challenging magnetic Ap targets for direct field detection and characterisation. Low-resolution investigations \citep{kochukhov:2006,hubrig:2006b} did not detect the field whereas the high-resolution spectropolarimetric study by \citet{auriere:2007} reported only two definite Zeeman detections out of 14 observations. Polarisation signatures are evidently very weak for \her, requiring one to use a multi-line technique to enhance the Zeeman signal. Here we rely on the least-squares deconvolution \citep[LSD,][]{donati:1997} method as implemented in the {\sc iLSD} code by \citet*{kochukhov:2010a}.

To derive the Stokes $I$ and $V$ LSD profiles we extracted a custom line list from VALD with the abundances and atmospheric parameters derived in Sect.~\ref{sec:params}. Spectral regions contaminated by telluric absorption or affected by broad wings of the hydrogen Balmer lines were excluded. Metal lines deeper than 5 per cent of the continuum prior to any macroscopic broadening were selected for the LSD line mask, yielding 2720 line in total. LSD profiles were calculated on the velocity grid spanning a range of $\pm300$~\kms\ using a step of 2~\kms. The LSD line weights were normalised using the mean wavelength $\lambda_0=500$~nm and the effective Land\'e factor $z_0=1.2$ (see \citealt{kochukhov:2010a} for details).

\begin{table}
\caption{Physical parameters of \her. \label{tab:params}}
\begin{tabular}{ll}
\hline
Parameter & Value \\
\hline
$T_{\rm eff}$ (K) & $9400\pm200$ \\
$\log g$ (cm\,s$^{-2}$) & $3.55\pm0.05$ \\
$R$ ($R_\odot$) & $4.3\pm0.1$ \\
$\log L$ ($L_\odot$) & $2.113\pm0.042$ \\
$M$ ($M_\odot$) & 2.71--2.97 \\
Age (Myr) & 351--469 \\
$P_{\rm rot}$ (d) & $4.116476\pm0.000022$ \\
$v_{\rm e}\sin{i}$ (\kms) & $35.5\pm0.5$ \\
$i$ (\degr) & $42.2\pm1.4$ \\
\hline
\end{tabular}
\end{table}

This application of the LSD procedure to Stokes $V$ observations of \her\ provided profiles with a median $S/N$ of 15500, corresponding to a factor 30 gain relative to the original spectra. This allowed us to detect Zeeman signatures at many rotational phases. We used the false alarm probability (FAP) detection diagnostic (\citealt*{donati:1992}, \citealt{donati:1997}) to quantify these detections. Following well-established criteria, we consider a detection to be definite if FAP\,$<$\,$10^{-5}$, marginal detection if $10^{-5}$\,$<$\,FAP\,$<$\,$10^{-3}$, and non-detection for FAP\,$>$\,$10^{-3}$. This detection diagnostic is reported in the sixth column of Table~\ref{tab:obs} for all 46 Narval spectra. Considering these original observations without averaging close rotational phases, we get 27 definite detections, 4 marginal detections, and 15 non-detections. Applying the same criteria to the LSD profiles calculated for the diagnostic null spectra yields only non-detections. If 11 pairs of consecutive observations obtained on the same observing nights are averaged, the detection statistics changes to 22 definite detections, one marginal detection, and 12 non-detections out of 35 observations. There are still no significant signals in the null profiles.

In addition to calculating LSD profiles using all available metal lines, we derived several sets of mean profiles for individual chemical elements. Although none of the resulting LSD Stokes $V$ profiles yield usable magnetic signatures, the corresponding LSD Stokes $I$ spectra provide a distilled high $S/N$ observable useful for mapping chemical spot distributions over the stellar surface. Three chemical elements -- Fe (1232 lines), Cr (979 lines), and Ti (122 lines) -- have sufficiently large number of absorption features in the spectrum of \her\ for a meaningful application of the LSD method. Average profiles of these chemical elements were derived using the multi-profile mode of the {\sc iLSD} code, which allowed us to account the background blending by other elements.

\subsubsection{Longitudinal magnetic field}

The mean longitudinal magnetic field, \bz, is a widely used measure of the disk-averaged line of sight component of stellar magnetic geometries \citep[e.g.][]{mathys:1991}. This magnetic characteristic can be determined from the Stokes $I$ and $V$ LSD profiles by calculating the first moment of the $V$ profile about the line's centre of gravity and normalising it by the equivalent width of the $I$ profile \citep{wade:2000,kochukhov:2010a}. The results of these calculations for \her\ using a $\pm45$~\kms\ integration window are presented in the seventh column of Table~\ref{tab:obs}.

\begin{figure}
\centering
\figps{\hsize}{0}{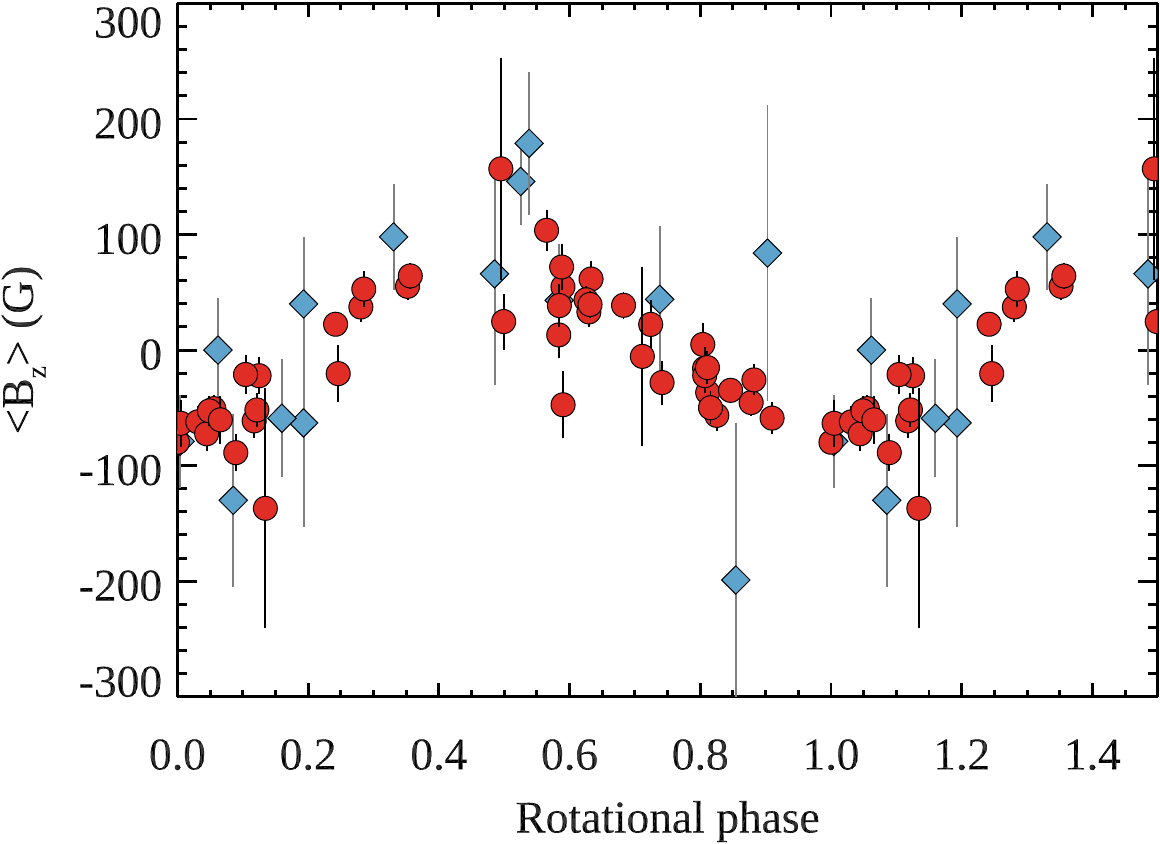}
\caption{Mean longitudinal magnetic field of \her\ as a function of the rotational phase. Circles correspond to the measurements obtained in this study from Narval spectropolarimetric observations whereas diamonds show the data from \citet{auriere:2007}.}
\label{fig:bz}
\end{figure}

This mean longitudinal magnetic field analysis shows that \bz\ of \her\ varies approximately sinusoidally between $-100$ and $+100$~G. Figure~\ref{fig:bz} compares our measurements with the outcome of previous magnetic analysis of this target by \citet{auriere:2007} using the ephemeris derived in Sect.~\ref{sec:period}. It is evident that Narval spectra enable a major improvement in the \bz\ measurement precision. Specifically, our \bz\ data set has a median error bar of 15~G and yields $>$3$\sigma$ longitudinal field detections for 24 out of 46 observations. This can be contrasted with a single $>$3$\sigma$ measurement in \citet{auriere:2007}. At the same time, derivation of the mean longitudinal field from the LSD profiles corresponding to the diagnostic null spectra yields no instances of \bz\ measurements exceeding 2.3$\sigma$, which confirms that out longitudinal field uncertainties are realistic.

\subsection{Zeeman Doppler imaging}
\label{sec:zdi}

\subsubsection{Methodology and codes}
\label{sec:method}

A variety of inversion codes and tomographic reconstruction methods was employed in our investigation of the surface structure of \her. Most of the inversions were carried out with the {\sc InversLSD} code, introduced by \citet{kochukhov:2014} and subsequently used in several studies of Ap stars \citep{kochukhov:2017a,kochukhov:2019,kochukhov:2022}. This software enables reconstruction of a vector magnetic field distribution simultaneously with one scalar parameter map representing local continuum brightness, line strength, temperature or element abundance. The input observational data for this inversion method comprises time series of LSD profiles in one or several Stokes parameters. {\sc InversLSD} uses pre-computed tables of local Stokes parameter profiles, which can be calculated under different assumptions and, in particular, treating the polarised radiative transfer problem with different levels of sophistication.

Considering the weakness of the observed circular polarisation signatures in the spectrum of \her, we have to rely on modelling the average metal-line LSD Stokes $V$ profiles for the reconstruction of magnetic field topology. The corresponding Stokes $I$ LSD profiles, also obtained by averaging over all metal lines, show strong rotational modulation. This variability corresponds to a weighted average of variabilities of all chemical elements contributing to the LSD line mask (with the largest contributions coming from Fe, Cr, and Ti) and cannot be ascribed to an inhomogeneous surface distribution of one particular element. Nevertheless, it is essential to reproduce this mean metal-line Stokes $I$ LSD profile variability in detail since it affects the Stokes $V$ observable. We model this variability using a non-uniform line strength (or equivalent width) surface distribution implemented in the context of the Unno-Rachkovsky analytical solution of the polarised radiative transfer problem in the Milne-Eddington atmosphere (hereafter ME-UR, \citealt{polarization:2004}). This treatment assumes that the LSD profile behaves as a single spectral line split as a Zeeman triplet with the central wavelength and Land\'e factor equal to the mean values of these parameters in the LSD mask. This approach to calculating the local Stokes parameter profiles is commonly employed by the ZDI studies of cool active stars \citep[e.g.][]{morin:2008,kochukhov:2017c,kochukhov:2020a,lehtinen:2022}, except that here we use the local line strength instead of continuum brightness to model the star spot signatures in Stokes $I$ spectra.

A different treatment is required to derive physically meaningful surface maps of individual chemical elements from their LSD profiles. As demonstrated by \citet{kochukhov:2010a}, the response of an LSD profile of particular chemical element to the abundance variation of that element cannot be reproduced by calculations assuming that LSD profile behaves as a single fiducial spectral line. Instead, theoretical LSD profiles correctly reproducing this variation can be computed by applying least-squares deconvolution to grids of detailed radiative transfer calculations for the entire wavelength domain covered by observations. We refer to our previous ZDI studies of Ap stars for an in-depth discussion of the technical implementation and numerical details of this approach \citep{kochukhov:2014,kochukhov:2017a,kochukhov:2019,kochukhov:2022}. Here this methodology is applied to deriving surface distributions of Fe, Cr, and Ti from the Stokes $I$ LSD profiles of these elements using {\sc InversLSD}, this time with a fixed magnetic field determined from the mean metal-line LSD profiles.

Further chemical elements besides those mentioned above show interesting rotational variability in \her. In particular, modulation of the \ion{O}{i} 777.19-777.54~nm triplet is noteworthy. This variability is not amenable to LSD modelling due to the small number of diagnostic lines. In this case, it is more appropriate to apply a conventional DI technique optimised for modelling individual spectral lines. We accomplished this using the {\sc Invers10} ZDI code \citep{piskunov:2002a,kochukhov:2002c}, which has been employed by many previous studies of strong-field Ap stars (e.g. \citealt{kochukhov:2010}; \citealt*{silvester:2014,silvester:2015}; \citealt{rusomarov:2016}, \citealt*{rusomarov:2018}). 

Additional complication associated with the spectrum synthesis modelling of the near-infrared \ion{O}{i} triplet is a significant departure of the formation of these lines in the atmospheres of A and B-type stars from LTE (e.g. \citealt{takeda:1997a}; \citealt*{sitnova:2013}; \citealt{takeda:2016}). Aiming to obtain an accurate O abundance map, we treated departures from LTE with dedicated statistical equilibrium calculations for oxygen atoms in the atmosphere of 45 Her. For this, we used the non-LTE (NLTE) code {\sc Balder} \citep{amarsi:2018}, which is a version of {\sc Multi3D} \citep{leenaarts:2009} with various updates in particular to the equation of state and background opacities.  The model atom was the same as that described in \citet{amarsi:2018a}, and uses B-spline R-matrix (BSR) data for electron-impact excitation  \citep{zatsarinny:2006,tayal:2016}. This model atom was updated with Stark broadening parameters taken from the Kurucz online database \citep{kurucz:1995}. The calculations were carried out for a range of O concentrations spanning 6 dex in logarithmic abundance, using a step of 0.2 dex, under the trace-element assumption. {\sc Invers10} was modified to read and interpolate within the resulting tables of NLTE departure coefficients. These data were incorporated in the polarised radiative transfer calculations by modifying the oxygen line opacity and source function according to well-known relations \citep[e.g.][]{piskunov:2017}. Fixed magnetic field derived with {\sc InversLSD} from the mean metal-line LSD profiles was included in {\sc Invers10} NLTE O abundance mapping.

In all three tomographic mapping approaches used in this study the global stellar magnetic field was represented with the help of a spherical harmonic expansion into poloidal and toroidal components \citep{kochukhov:2014}. This expansion included harmonic modes up to $\ell_{\rm max}=10$ and was constrained with a harmonic penalty function \citep{morin:2008,kochukhov:2014} to minimise contribution of higher-order harmonic terms not justified by the observational data. The line strength and chemical abundance maps were prescribed on a 1876-element surface grid and regularised by minimising the local surface gradients \citep[Tikhonov regularisation, e.g.][]{piskunov:2002a}. The optimal magnetic field and line strength / abundance regularisation parameters, corresponding to a compromise between obtaining a satisfactory fit to observations while avoiding spurious small-scale features, were established with the iterative regularisation stepping procedure described by \citet{kochukhov:2017}.

\subsubsection{Rotational period and differential rotation}
\label{sec:period}

We started the Doppler imaging analysis of \her\ by refining the stellar \vsini, determining rotational period, and searching for a differential rotation. To this end, we made use of the Stokes $I$ LSD profiles derived using all metal lines. Consecutive observations obtained on the same nights were averaged, resulting in a data set comprising 35 spectra. These observations were modelled with a non-uniform surface line strength distribution under the ME-UR local line profile approach described above. Given the weakness of the stellar magnetic field, it was ignored in this analysis.

\begin{figure}
\centering
\figps{0.9\hsize}{0}{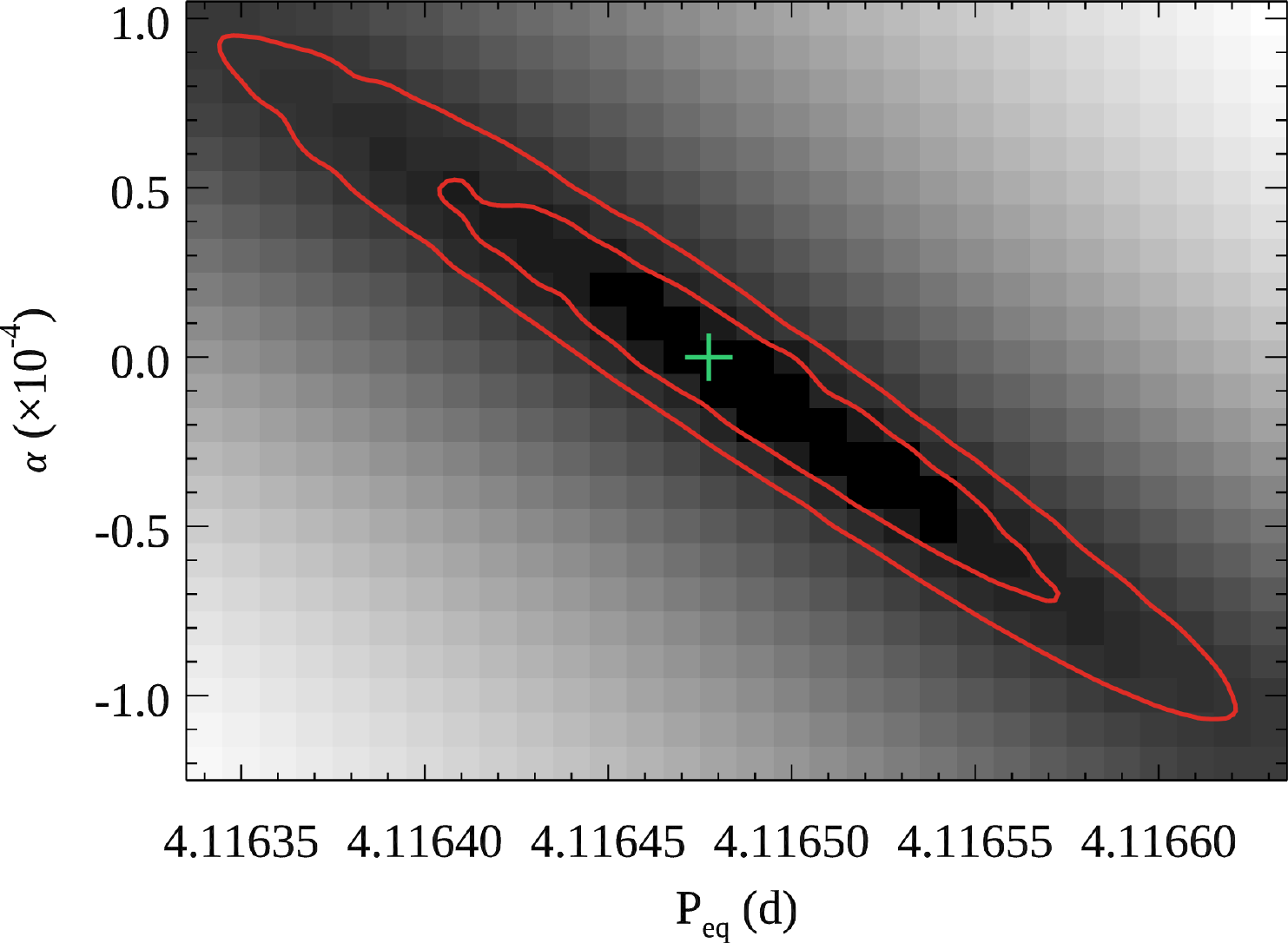}
\caption{Relative $\chi^2$ of the fit to Stokes $I$ LSD profiles as a function of equatorial rotational period $P_{\rm eq}$ and differential rotation parameter $\alpha$. The solid lines show the 68 and 90 per cent confidence intervals. The cross corresponds to the best-fitting solid-body rotational period.}
\label{fig:difrot}
\end{figure}

Inversions were repeated for different rotational periods around the previously known value $P_{\rm rot}=4.1164\pm0.0001$~d \citep{hatzes:1991a,paunzen:2021b}. Simultaneously, a range of differential rotation behaviours was explored assuming that the stellar surface is sheared according to the simplified solar law commonly adopted by DI and ZDI studies (e.g. \citealt{donati:1999}; \citealt*{petit:2002})
\begin{equation}
\Omega(l) = \Omega_{\rm eq} (1 - \alpha \sin^2 l ),
\end{equation}
where $\Omega(l)$ is the latitudinal dependence of the rotation rate, $\Omega_{\rm eq}=1/P_{\rm eq}$ is the equatorial rotation rate, and $\alpha=\Delta\Omega/\Omega_{\rm eq}$ is the normalised difference between rotation rates of the equator and pole. Using this approach, the $\chi^2$ of {\sc InversLSD} fit to the LSD profiles was studied as a function of $P_{\rm eq}$ and $\alpha$ with a resolution of $10^{-5}$ in both of these parameters. This analysis, illustrated in Fig.~\ref{fig:difrot}, revealed no evidence of either a solar-like or an anti-solar differential rotation with the upper limits $7\times10^{-5}$ (68 per cent confidence) and $11\times10^{-5}$ (90 per cent confidence) derived for the $|\alpha|$ parameter. In the limit of a solid-body rotation, we found $P_{\rm rot}=P_{\rm eq}=4.116476\pm0.000022$~d. Adopting the mid-exposure time of the first Narval observation as the reference time yields the ephemeris
\begin{equation}
HJD = 2454173.71327 + 4.116476\times E,
\end{equation}
which we use throughout this study.

At this stage of the DI analysis we also optimised the stellar projected rotational velocity. The best fit to observations was achieved using \vsini\,=\,$35.5\pm0.5$~\kms, in good agreement with previous findings ($35.0\pm0.5$~\kms\ in \citealt{hatzes:1991a}, 35~\kms\ in \citealt{auriere:2007}). Combining $P_{\rm rot}$ and \vsini\ derived here with $R=4.3\pm0.1$~$R_\odot$ determined in Sect.~\ref{sec:params} one can estimate inclination of the stellar rotational axis, $i=42.2\degr\pm1.4\degr$, using the relation between stellar radius, rotational period, and equatorial rotational velocity \citep{prsa:2016a}. This combination of $P_{\rm rot}$, \vsini, and $i$ was employed for all magnetic and abundance inversions reported below. The final comparison of the observed metal-line Stokes $I$ LSD profiles with the {\sc InversLSD} fit is shown in the left panel of Fig.~\ref{fig:prfiv}.

\subsubsection{Magnetic field topology}
\label{sec:mag}

\begin{figure}
\centering
\figps{\hsize}{0}{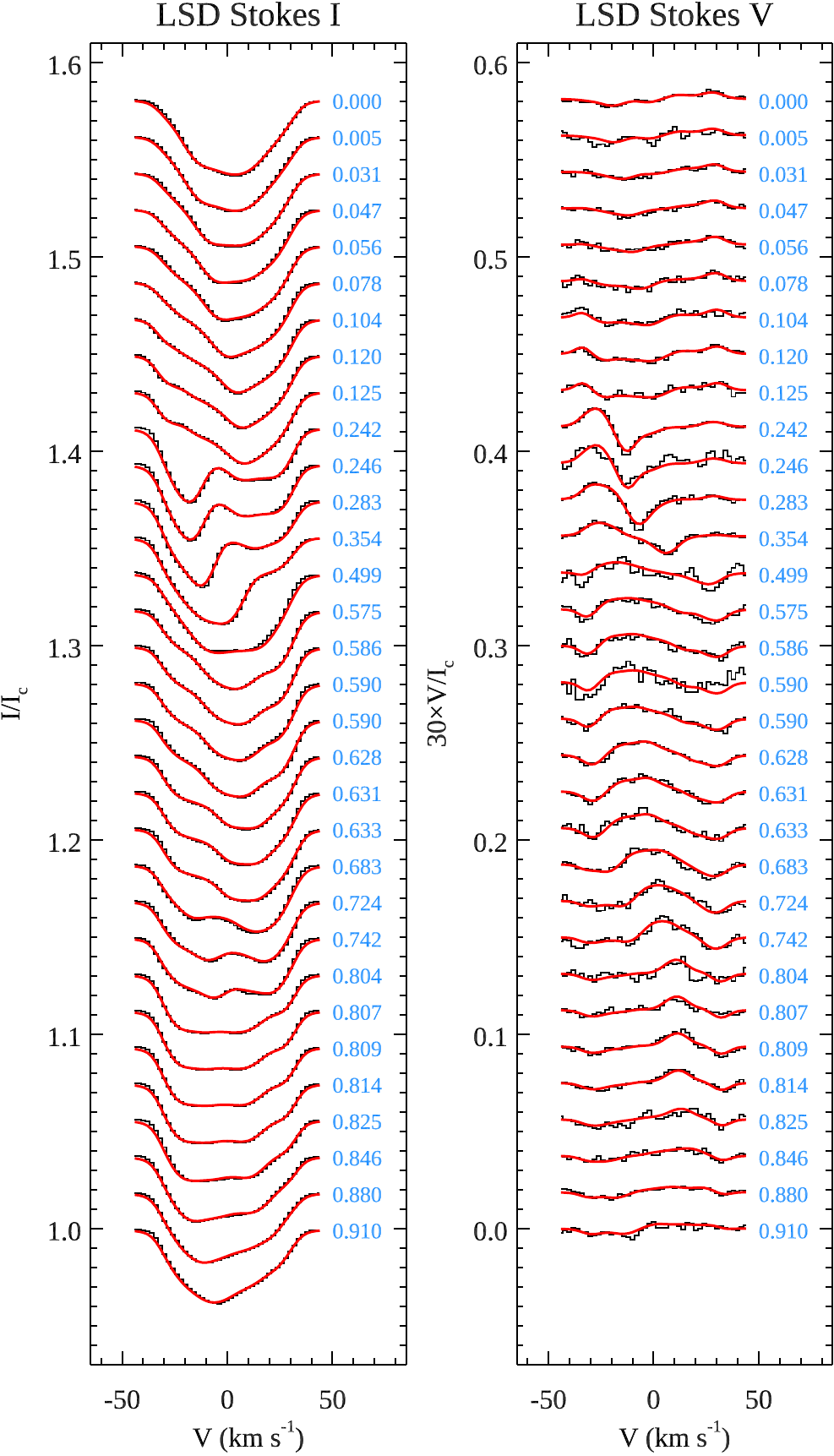}
\caption{Comparison of the observed (black histograms) and model (red solid lines) LSD Narval Stokes $I$ and $V$ profiles corresponding to the full metal line mask. Profiles are shifted vertically and arranged according to the rotational phase, which is indicated to the right of each profile.}
\label{fig:prfiv}
\end{figure}

\begin{figure*}
\centering
\figps{0.9\hsize}{270}{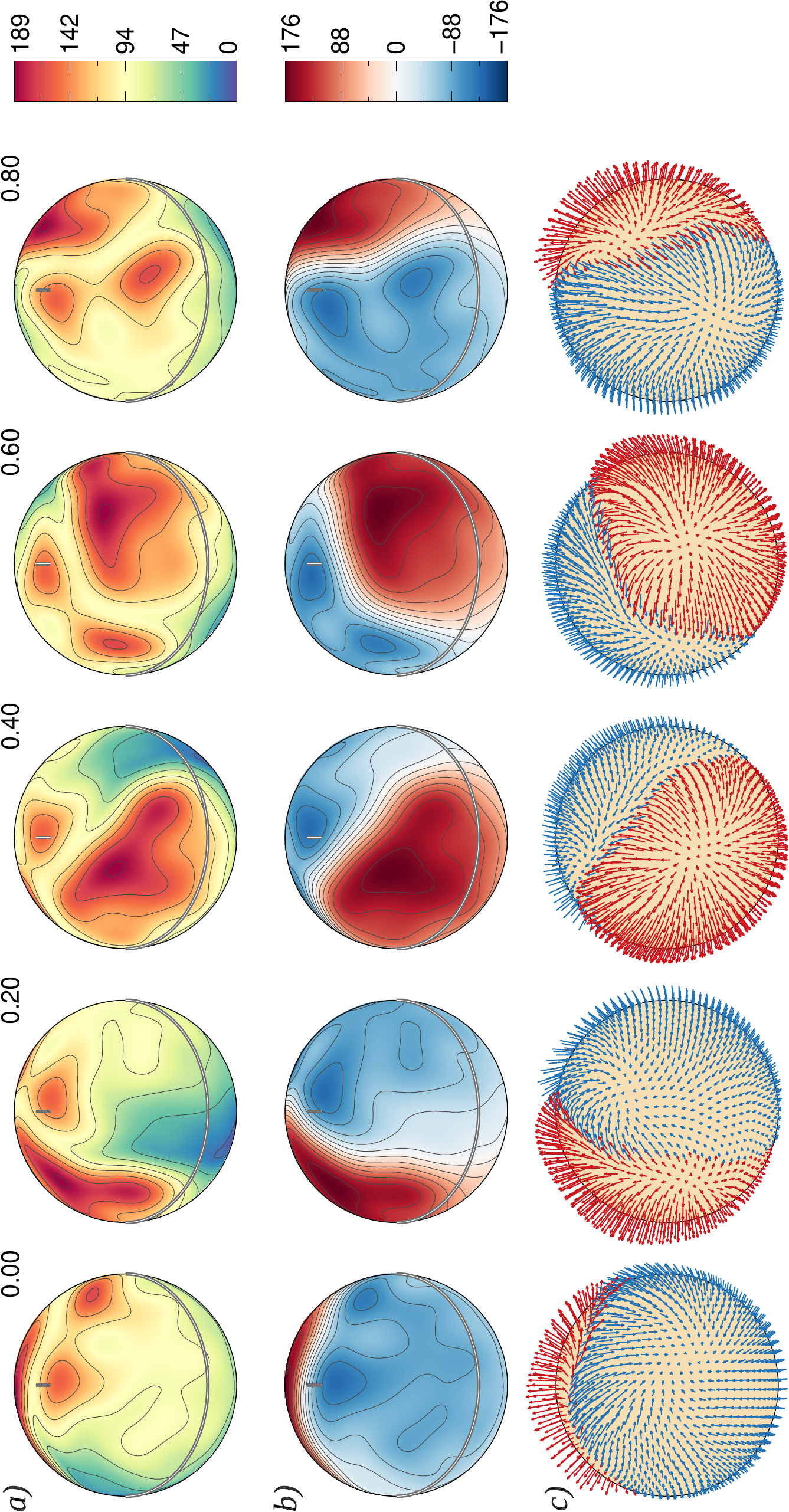}
\caption{Magnetic field topology of \her\ derived with ZDI analysis from metal-line Stokes $V$ LSD profiles. The three rows of spherical plots show maps of the magnetic field strength (a), the radial field component (b), and the magnetic field vector (c). The star is shown at the inclination $i=42\degr$ for five rotational phases, which are indicated next to each column. The contours over spherical maps in panels (a) and (b) are plotted with a step of 25 G. The stellar equator and visible rotational pole are indicated with double lines. The side colour bars give the field strength in G.}
\label{fig:fld}
\end{figure*}

Magnetic field geometry of \her\ was derived with the ME-UR {\sc InversLSD} modelling of the mean metal-line Stokes $V$ profiles after constraining the line strength map as described in the previous section. Three observations with low $S/N$ were excluded from this analysis and spectra obtained on the same nights were averaged, resulting in a data set comprising 32 observations. The comparison between the observed and best-fitting model circular polarisation spectra is shown in the right panel of Fig.~\ref{fig:prfiv}. Spherical renderings of the vector magnetic field map corresponding to this fit are presented in Fig.~\ref{fig:fld}. 

We find that the global magnetic field of \her\ is roughly dipolar, featuring a large inclination of the dipolar axis relative to the stellar axis of rotation. At the same time, this dipolar topology is significantly distorted, particularly if one examines the field modulus map (Fig.~\ref{fig:fld}a). There is a noticeable asymmetry between the negative and positive magnetic poles, with the latter one exhibiting stronger field. The local field modulus peaks at 190~G in a spot located within the positive radial field region but does not exceed 166~G in the part of the surface covered by negative field. In both regions the structures with the strongest field appear to be offset towards the magnetic equator. The surface-averaged field strength is 77~G.

According to the harmonic power spectrum of the derived magnetic field distribution, the field of \her\ can be characterised (see Table~\ref{tab:mag}) as mostly dipolar (74.6 per cent of the magnetic field energy), poloidal (86.4 per cent) and non-axisymmetric (95.5 per cent). The equivalent dipolar field strength calculated from $\ell=1$ harmonic coefficients is 119~G. Contribution of $\ell\ge2$ modes responsible for the apparent distortion of global dipolar field in Fig.~\ref{fig:fld} is significant. An attempt to fit observations using a purely dipolar, poloidal field increased the standard deviation between the observed and model Stokes $V$ spectra by 40 per cent giving an unsatisfactory fit. On the other hand, excluding toroidal components from $\ell_{\rm max}=10$ harmonic expansion yields only a marginal decrease of the fit quality while maintaining similar surface magnetic field structure. This suggests that the toroidal part of 45\,Her's field topology is poorly constrained by the available observations and therefore detection of this field component is rather uncertain.

\begin{table}
\caption{Magnetic field characteristics of \her. \label{tab:mag}}
\begin{tabular}{ll}
\hline
Parameter & Value \\
\hline
$E_{\rm pol}$ & 86.4\% \\
$E(|m|>0)$ & 95.5\% \\
$E(\ell=1)$   & 74.6\% \\
$E(\ell=2)$   & 15.3\% \\
$E(\ell=3)$   & 4.3\% \\
$E(\ell\ge4)$  & 5.8\% \\
\hline
$\langle B \rangle$ & 77.5 G \\
$\langle |B_{\rm r}| \rangle$ & 66.1 G \\
$B_{\rm d}$ & 118.8 G \\
\hline
\end{tabular}
\end{table}

\subsubsection{Chemical spot distributions}

Abundance distributions of Ti, Cr, and Fe were derived individually from the Stokes $I$ LSD profiles of these elements using the {\sc InversLSD} code and following the methodology described in Sect.~\ref{sec:method}. The magnetic field was kept fixed in these inversions according to the results of Sect.~\ref{sec:mag}. The resulting comparison of the observed and model LSD profiles is illustrated in Figs~\ref{fig:prfti}--\ref{fig:prffe} for Narval data. The same analysis procedure was applied independently to the LSD spectra derived from CES observations. The corresponding model fits are presented in Figs~\ref{fig:prftiT}--\ref{fig:prffeT}. For oxygen, we instead used the {\sc Invers10} code modified to account for NLTE effects and directly modelled the near-infrared \ion{O}{i} triplet in Narval spectra as illustrated in Fig.~\ref{fig:prfo}. This figure also shows the impact of ignoring departures from LTE, which leads to a major underestimation of the O line absorption for a given oxygen abundance.

\begin{figure*}
\centering
\figps{0.9\hsize}{270}{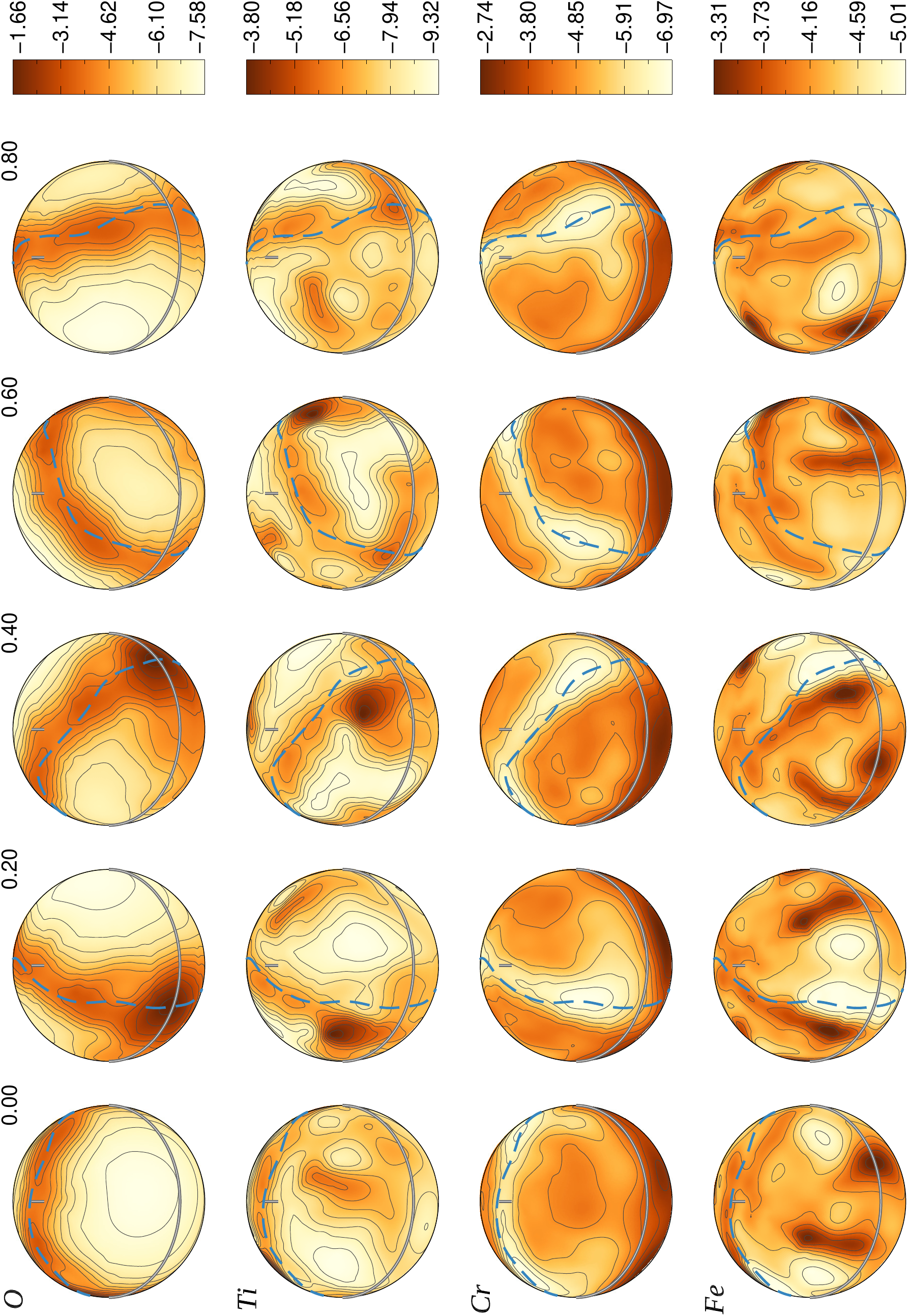}
\caption{Surface abundance distributions of O, Ti, Cr, and Fe obtained with DI modelling of Stokes $I$ Narval spectra obtained during 2018. The star is shown at the inclination $i=42\degr$ for five rotational phases, which are indicated next to each column. The contours over spherical maps are plotted with a step of 0.25 dex for Fe and 0.50 dex for other elements. The fractional element abundances are given by the side bars in the $\log (N_{\rm el}/N_{\rm tot})$ units. The double lines correspond to the stellar equator and rotational pole. The thick dashed line shows where the radial magnetic field changes sign according to the magnetic map in Fig~\ref{fig:fld}.}
\label{fig:abn}
\end{figure*}

\begin{figure*}
\centering
\figps{0.9\hsize}{270}{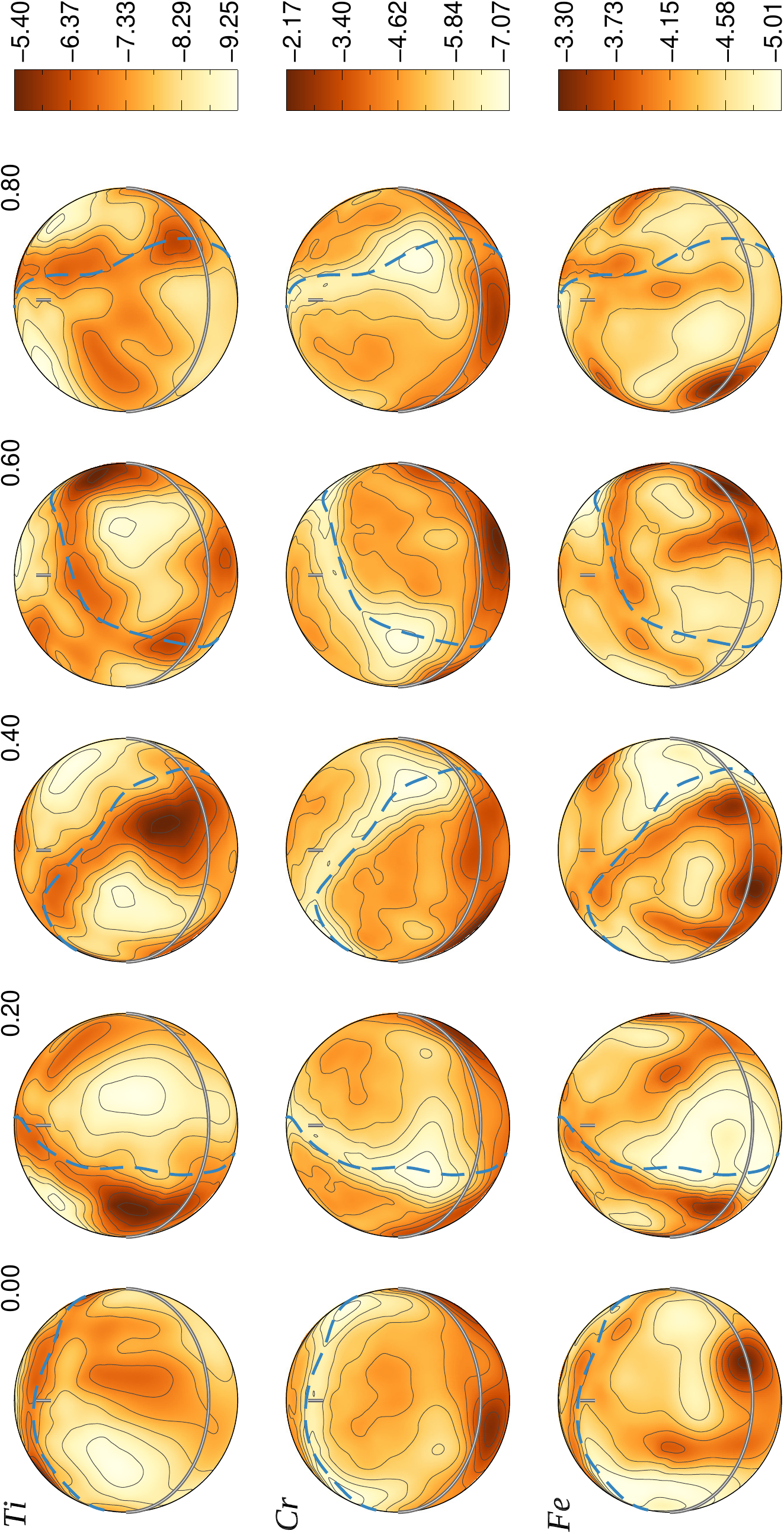}
\caption{Same as Fig.~\ref{fig:abn} but for the surface abundance distributions of Ti, Cr, and Fe obtained from CES observations obtained between 2014 and 2018.}
\label{fig:abnT}
\end{figure*}

The orthographic projection of the O, Ti, Cr, and Fe spot distributions obtained from Narval observations is presented in Fig.~\ref{fig:abn}, where the star is shown at five rotational phases and at the actual inclination angle as seen by the observer. The same plot for Ti, Cr, and Fe spot maps derived from CES data is given in Fig.~\ref{fig:abnT}. A comparison between these two sets of chemical abundance maps will be discussed in Sect.~\ref{sec:stability}.

For all four studied chemical elements we found large-scale, high-contrast abundance inhomogeneities spanning several dex in logarithmic element concentration. The O map exhibits a ring-like structure with a contrast reaching nearly 6~dex. Additionally, there is a spot characterised by an extreme concentration of O at one of the two intersections of this ring with the stellar rotational equator. This asymmetry of the reconstructed O distribution is not an inversion artefact since the O triplet lines clearly appear stronger and wider at phase $\approx$\,0.35 compared to phase $\approx$\,0.80 when the two sides of the overabundance ring cross the disk centre. The O map reconstructed in LTE is qualitatively similar but has much higher local abundance values.

The Ti distribution exhibits the second highest surface contrast of around 5~dex. The behaviour of this element resembles that of O, although in addition to a ring-like structure it shows a number of smaller spots and a more articulated high-contrast feature at phase 0.3. The location of this spot is close to the O overabundance region at the stellar equator.

In contrast to O and Ti, the distribution of Cr is dominated by a ring of relative underabundance spanning the visible hemisphere. For this element our inversion procedure also recovers major overabundance features below the rotational equator. However, as seen from Fig.~\ref{fig:abn}, these parts of the stellar surface are always visible near the limb and contribute little to the disk-integrated spectra. Reliability of these surface features is therefore low.

Finally, Fe shows the lowest abundance contrast (about 1.5 dex) and the most complex surface distribution among all four studied chemical elements. The iron map exhibits at least five latitudinally-extended overabundance spots or arcs. This complexity can be traced back to the rotational modulation of Fe lines (e.g. Fig.~\ref{fig:prffe}), which shows many small-amplitude distortions travelling across rotationally-broadened profile compared to a couple of dominant spot features visible in the lines of O, Ti, and Cr.

\section{Summary and discussion}
\label{sec:disc}

In this study we presented a detailed investigation of the physical parameters, binarity, magnetic field and surface chemical spot distributions of the magnetic Ap star \her. This star was found to be a wide, single-lined spectroscopic binary with $P_{\rm orb}=99.7$~d and a radial velocity semi-amplitude of $K_1=4.7$~\kms. Its companion is a late-type, low-mass star that makes no detectable contribution to the composite spectra. We revised atmospheric parameters of \her\ based on model atmosphere fitting of the H$\gamma$ line profile and several spectrophotometric observations, including recently released Gaia DR3 spectrophotometry. These data are best reproduced with $T_{\rm eff}=9400$~K, $\log g=3.55$ and $R=4.3$~$R_\odot$. Abundance analysis based on the mean spectra revealed a typical chemical pattern of an evolved magnetic Ap star (\citealt{kochukhov:2006b}; \citealt*{bailey:2014a}): an underabundance of light elements, Ca and Ti,  approximately normal concentration of Mg and Si, and large overabundance of Fe-peak elements without a detectable enhancement of rare-earths. Zeeman circular polarisation signatures were repeatedly detected in high-resolution spectropolarimetric observations of \her, indicating a reversing longitudinal magnetic field with a strength below 100~G. We used the LSD circular polarisation profiles to reconstruct the surface magnetic field topology of \her\ with the help of the ZDI technique. The spatially-resolved surface magnetic field strength of that star peaks at 190~G with the mean field modulus of 77~G. The stellar field geometry is dominated by the poloidal, non-axisymmetric components and is roughly dipolar, although contributions of higher order components (quadrupole and above) is non-negligible. Applying Doppler imaging analysis to LSD intensity profiles derived from all metal lines, we demonstrated the absence of a latitudinal differential rotation with an upper limit of $| \alpha |=|\Delta\Omega | / \Omega \approx 10^{-4}$. We then used the LSD intensity profiles of Ti, Cr, and Fe to reconstruct spot distributions of these elements. This analysis was repeated for two independent data sets obtained at different observatories. Finally, we used the higher resolution better $S/N$ Narval data set to recover the oxygen surface abundance map from the spectacular variation of the near-infrared \ion{O}{i} triplet based on the spectrum synthesis calculations which incorporated NLTE line formation of that element.

\subsection{Magnetic field of \her\ compared to other Ap stars}

The distinguishing feature of \her\ is its remarkably weak surface field for a magnetic Ap star. These objects are known to have fields with dipolar components ranging from several hundred G up to $\sim$\,30~kG \citep[e.g.][]{auriere:2007,sikora:2019a}. However, the accuracy and reliability of these field strength estimates, especially for weaker fields, depend on the quality of observations and degree of detail with which these fields are studied. To this end, \her\ has the lowest magnetic field strength among Ap stars modelled with ZDI \citep{kochukhov:2019} and the second lowest field among all Ap stars analysed with modern high-resolution spectropolarimetry. The only star with a weaker field, HD\,5550 \citep{alecian:2016}, is an atypical object in several respects. It is a short-period ($P_{\rm orb}=6.84$~d), double-lined spectroscopic binary. The field with a dipolar strength of $B_{\rm d}=65\pm20$~G is nearly aligned with the orbital axis and is found on the primary component, which rotates with the same period as its orbital motion and lacks significant spectral variations indicative of chemical spots. In contrast, \her\ is effectively a single object that exhibits all staple characteristics of magnetic Ap stars (non-solar chemical abundances, highly inclined distorted dipolar field, prominent rotational modulation due to chemical spots). This proves that the chemical peculiarity and surface structure formation phenomena associated with a fossil magnetic field in late-B and A-type stars extend down to dipolar field strengths of $\approx$\,120~G (corresponding to a surface-averaged field strength of $\approx$\,80~G). This is lower than thought previously but still noticeably higher than the $\approx$\,30~G dipolar field of the Am star $\gamma$~Gem \citep{blazere:2020} and orders of magnitude stronger than sub-G magnetic fields found on other Am and normal intermediate-mass stars \citep{petit:2010,petit:2011,blazere:2016}.

\begin{figure*}
\centering
\figps{\hsize}{90}{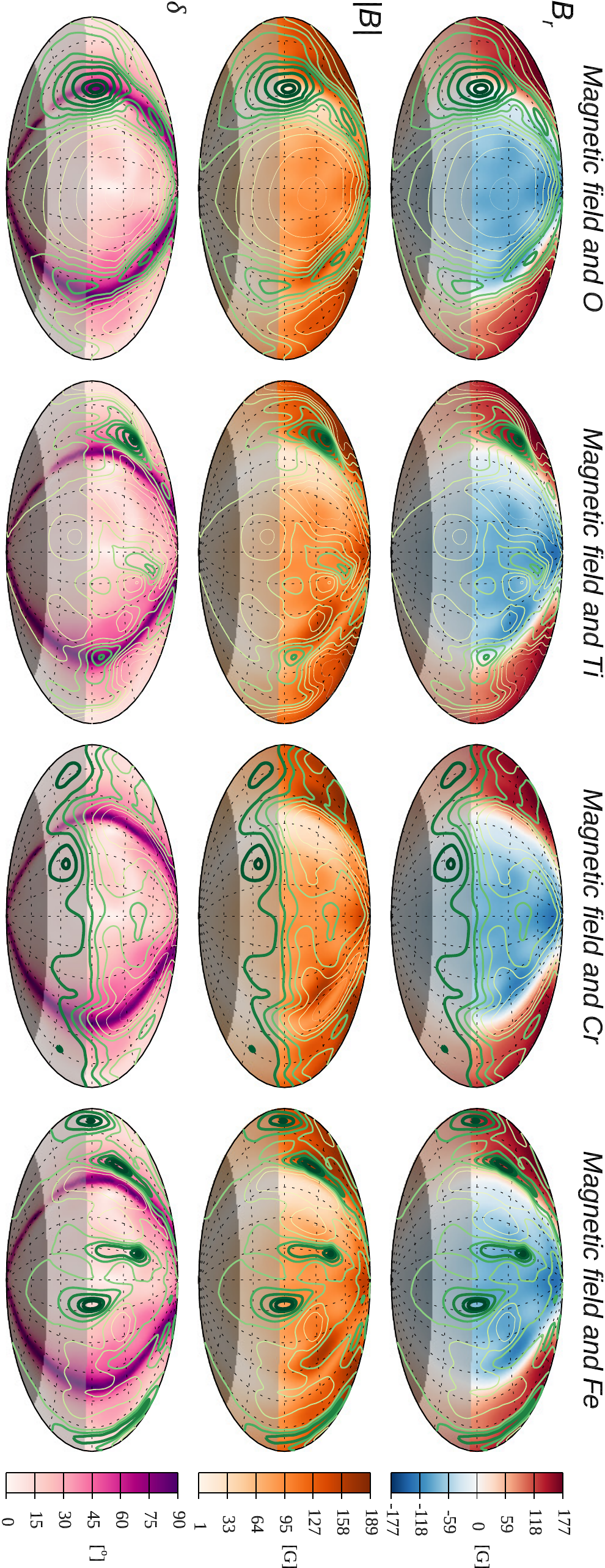}
\caption{Comparison of the magnetic field topology of \her\ with abundance distributions for O, Ti, Cr, and Fe based on Narval observations. The three rows of Hammer-Aitoff projection plots display in the background, from top to bottom, the radial magnetic field component, the field strength, and the local field inclination. The dotted lines show the longitude-latitude grid plotted with a 30\degr\ step and zero longitude in the centre. The shaded bottom part of the maps indicates latitudes that are completely invisible (dark grey) or have a relative visibility of less than 25 per cent (light grey). Chemical abundance distributions are shown with foreground contours plotted every 0.25 dex for Fe and 0.50 dex for other elements. The darker and thicker contours correspond to regions with a higher element abundance.}
\label{fig:ham}
\end{figure*}

\subsection{Stability of weak fossil magnetic fields}

Owing to the weakness of its magnetic field, \her\ is a key object in the discussion of the hydromagnetic stability of fossil fields in early-type stars and of the ensuing lower field strength threshold. Based on a sample of 24 weak-field magnetic Ap stars, \citet{auriere:2007} has established that these objects appear to possess dipolar fields stronger than 300~G with the median strength estimated to be $B_{\rm d}\approx1.4$~kG. The existence of this 300~G threshold field strength was confirmed by \citet{sikora:2019a}, who studied a volume-limited sample of 52 magnetic Ap stars with higher quality data. They concluded that the mean dipolar field strength of their sample of Ap stars was 2.6~kG and also could not find a single star with $B_{\rm d}$ unambiguously below the 300~G threshold.

To explain the apparent lower limit of Ap-star fields, \citet{auriere:2007} put forward a hypothesis that the global fossil field is subject to a Taylor-type instability \citep[e.g.][]{spruit:1999} and cannot survive below a certain minimum field strength $B_{\rm c}$, which depends on the atmospheric equipartition field $B_{\rm eq}$, the stellar rotational period, radius and effective temperature,
\begin{equation}
B_{\rm c} = 2 B_{\rm eq} \left( \frac{P_{\rm rot}}{5 \mathrm{d}} \right)^{-1} \left( \frac{R}{3 R_\odot} \right) \left( \frac{T_{\rm eff}}{10^4 \mathrm{K}}\right)^{-1/2}.
\label{eq:bc}
\end{equation}
All stars studied by \citet{auriere:2007} and \citet{sikora:2019a} satisfied the condition $B_{\rm d} \ge B_{\rm c}$ but this is not the case for \her. Its equipartition field is $B_{\rm eq}=131$~G at $\tau_{500}=2/3$ for the $T_{\rm eff}=9400$~K, $\log g =3.55$ model atmosphere calculated in our study. Inserting other relevant stellar parameters in Eq.~(\ref{eq:bc}) yields $B_{\rm c}=471$~G. This is about a factor of 4 higher than the observed dipolar field strength of 119~G. Thus, the field of \her\ is clearly incompatible with the theory underlying the commonly discussed $\approx$\,300~G dipolar strength limit. Similar conclusions can be reached for the primary component of HD\,5550 \citep{alecian:2016} with the already mentioned caveat that the latter is part of a close binary system, which is likely to be relevant for both the origin and long-term survival of the fossil field. In contrast, a wide orbit and low mass of 45~Her's binary companion allows one to treat the primary as effectively a single star in the context of the field stability discussion. 

Another type of field stability assessment is commonly carried out in the context of 1D modelling of subsurface convection zones in the intermediate- and high-mass stars \citep[e.g.][]{cantiello:2009,cantiello:2011,cantiello:2019}. This treatment explores stability of the hydrogen and helium ionisation zones in lower mass stars and Fe-peak opacity zone in higher mass stars as a function of vertical magnetic field strength. The underlying idea is that the field strength has to exceed a certain limit, $B_{\rm c}$, to provide stability to convection. Fields weaker than $B_{\rm c}$ will be dragged, tangled and quickly destroyed by convective motions. Arguing along these lines, \citet{jermyn:2020} and \citet{farrell:2022} applied the convective stability criterium by \citet{macdonald:2019} (originally developed for O-star envelopes) to grids of 1D stellar structure models covering the mass interval relevant for magnetic Ap stars. According to the latest results by \citet{farrell:2022}, the fundamental parameters of \her\ require $B_{\rm c}$ of at least 1500~G. In this theory, the critical field corresponds to the surface-averaged radial magnetic field component. But, according to our ZDI results, the latter amounts to just 66~G for \her. Thus, there is a major contradiction between these theoretical predictions and observations. This tension points to a problem with the concept of a minimal vertical field required for stability to convection, likely stemming from an approximate 1D treatment of convection in the current stellar structure models. This also means that the far-reaching conclusions, for example on the separation between fossil and dynamo fields \citep{jermyn:2020} and on the initial magnetic field distribution of AB stars \citep{farrell:2022}, obtained using this theoretical framework have to be revised to account for the existence of stars like \her\ and HD\,5550~A.

\subsection{Relation between magnetic field and chemical spots}

We now take a closer look at the relation between magnetic field geometry and locations of chemical spots using the visualisation procedure introduced by \citet{kochukhov:2022}. Figure~\ref{fig:ham} shows the Hammer-Aitoff maps of the radial magnetic field component, the field modulus and the local field inclination relative to the surface normal simultaneously with the abundance distributions of O, Ti, Cr, and Fe. The parts of the stellar surface completely or partially invisible (as defined in \citealt{kochukhov:2022}) to the observer are indicated with shaded regions.

Starting with oxygen, we see a clear correlation of that element with the horizontal field regions. The O overabundance forms nearly a perfect ring at the magnetic equator, similar to the behaviour of this element in a few other magnetic Ap stars (\citealt*{rice:1997}; \citealt{kochukhov:2004e}). In \her\ this symmetry is, however, broken by an additional O spot located at one of the intersections of the magnetic and rotational equators. Turning to the field strength map, we can see that at this location the field intensity is systematically lower compared to the other side of the magnetic equator.

The distribution of Ti features a major spot that is located not far from the O spot, but is offset from the magnetic equator. There are two other, less prominent, Ti spots with the first one located at another intersection of the magnetic and rotational equators and the second one roughly coinciding with the positive magnetic pole. Titanium was rarely targeted by DI analyses of Ap stars and a few previous maps of this element reconstructed simultaneously with magnetic field did not show the tendency of concentration at the magnetic equator observed in \her\ (\citealt{luftinger:2010a}; \citealt*{silvester:2014a}; \citealt{rusomarov:2016}).

The surface distribution of Cr is dominated by a relative underabundance ring at the magnetic equator. This result agrees the earlier finding for \her\ and two other Ap stars by \citet{hatzes:1991a}. However, the trend for Cr to exhibit a depletion at the magnetic equator is rarely observed in Ap stars analysed with DI \citep[e.g.][]{kochukhov:2017a,kochukhov:2019,kochukhov:2022}.

The surface map of Fe reconstructed for \her\ is the most complex of all studied elements, comprising at least five separate Fe overabundance spots and three regions of prominent Fe depletion. One of the overabundance spots is co-located with the previously discussed major Ti spot in the vicinity of the weaker-field segment of the magnetic equator. At the same time, Fe is underabundant at this part of the magnetic equator itself, similar to Cr. Other features that can be discerned in the Fe surface map are pairs of spots located close to both negative and positive magnetic poles. Overall, the behaviour of Fe in \her\ shares the trend, seen in the DI results for other Ap stars \citep[e.g.][]{kochukhov:2004e,kochukhov:2014,kochukhov:2019}, for this element to exhibit a complex multi-spot distribution lacking a direct relation to the magnetic field geometry.

\subsection{Long-term stability of abundance distributions}
\label{sec:stability}

\begin{figure}
\centering
\figps{0.75\hsize}{90}{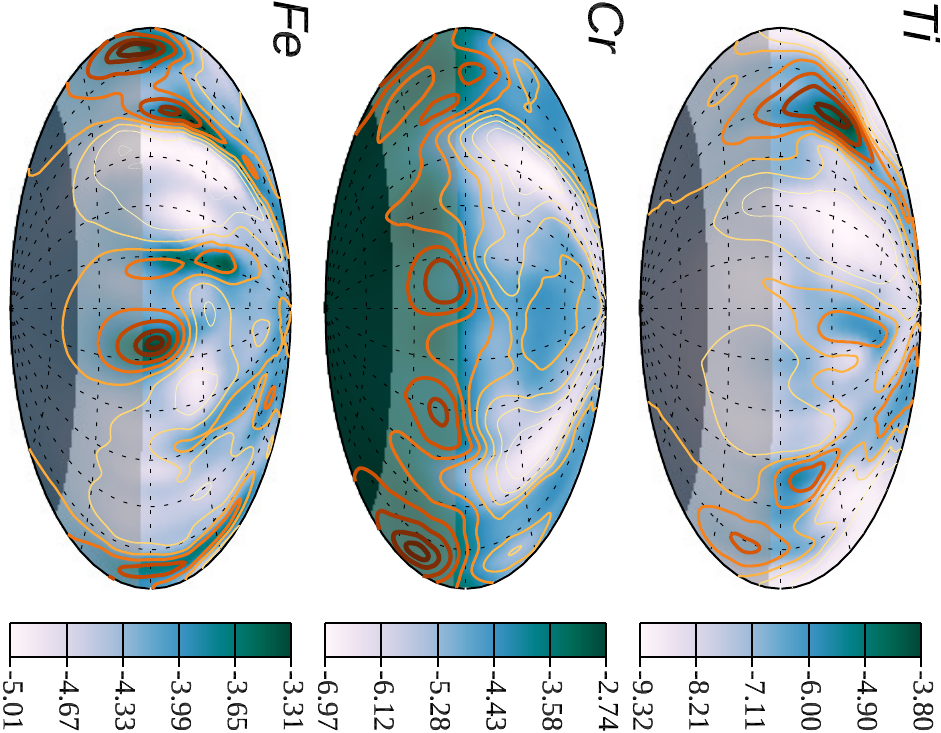}
\caption{Agreement between abundance distributions of Ti, Cr, and Fe reconstructed from independent spectroscopic data sets. The background Hammer-Aitoff plots show the abundance maps derived from the Narval spectra. The foreground contours correspond to the chemical spot distributions obtained from the CES observations. The contours are plotted every 0.25 dex for Fe and 0.50 dex for other elements, with darker colour and thicker lines indicating a higher element abundance.}
\label{fig:2abn}
\end{figure}

Independent abundance inversions using Narval and CES observations offer a unique opportunity to probe possible long-term evolution of chemical spots or, alternatively, assess uncertainties of DI reconstruction assuming a static surface distribution. It is known that, in the absence of strong global magnetic fields similar to those usually found in Ap stars, heavy-element spots on HgMn stars evolve on the time scales from years \citep{kochukhov:2007b} to months \citep{korhonen:2013}. Similar behaviour cannot be a priori excluded for \her\ considering the unusually weak magnetic field of this star.

All but two Narval observations were obtained in 2018 whereas most of the CES spectra were collected in 2014--2015. There is, therefore, a three year gap between the bulk of the two data sets. The spectral resolution, typical $S/N$, rotational phase coverage and reduction procedures applied to the Narval and CES data were all different. Nevertheless, the comparison of Ti, Cr, and Fe maps in Fig.~\ref{fig:2abn} shows an excellent agreement between the outcomes of the DI inversions using the two sets of spectra. In particular, every spot feature discussed above for the surface maps based on the Narval observations is reproduced in the maps inferred from the CES data. This even includes the most complex Fe map, for which the two independent inversions yield the same five overabundance spots. The only systematic difference between the two sets of maps, most clearly seen by comparing Figs~\ref{fig:abn} and \ref{fig:abnT}, is a smoother surface distribution derived using the CES observations, likely reflecting a lower resolution and $S/N$ of that data set. Thus, there is no evidence of any change in the surface maps of these three chemical elements between 2014--2015 and 2018. In this respect, \her\ behaves similar to other magnetic Ap stars and differently from non-magnetic HgMn stars.

Assuming static surface abundance distributions, we computed a mean absolute difference (weighted by the surface element visibility) of 0.32, 0.22, and 0.12 dex for Ti, Cr, and Fe, respectively. These numbers can be considered as conservative estimates of the local uncertainty of abundance reconstructions in this study. 

\section*{Acknowledgements}
OK acknowledges support by the Swedish Research Council (project 2019-03548) and the Royal Swedish Academy of Sciences.
AMA acknowledges support from the Swedish Research Council (project 2020-03940).
This work is based on observations obtained at the \textit{Bernard Lyot} Telescope (TBL, Pic du Midi, France) of the Midi-Pyr\'en\'ees Observatory, which is operated by the Institut National des Sciences de l'Univers of the Centre National de la Recherche Scientifique of France.
We thank to T\"UBITAK for a partial support in using RTT150 (Russian-Turkish 1.5-m telescope in Antalya) with the project numbers 13BRTT150-499 and 17BRTT150-1239.
The computations for this project were enabled by resources provided by the Swedish National Infrastructure for Computing (SNIC), partially funded by the Swedish Research Council through the grant agreement no. 2018-05973.
This research has made extensive use of the SIMBAD database, operated at CDS, Strasbourg, France.

\section*{Data availability}
The spectropolarimetric Narval observations underlying this article are available from the PolarBase archive (\url{http://polarbase.irap.omp.eu}). The CES observations will be shared on reasonable request to the corresponding author.


\begin{thebibliography}{}
\makeatletter
\relax
\def\mn@urlcharsother{\let\do\@makeother \do\$\do\&\do\#\do\^\do\_\do\%\do\~}
\def\mn@doi{\begingroup\mn@urlcharsother \@ifnextchar [ {\mn@doi@}
  {\mn@doi@[]}}
\def\mn@doi@[#1]#2{\def\@tempa{#1}\ifx\@tempa\@empty \href
  {http://dx.doi.org/#2} {doi:#2}\else \href {http://dx.doi.org/#2} {#1}\fi
  \endgroup}
\def\mn@eprint#1#2{\mn@eprint@#1:#2::\@nil}
\def\mn@eprint@arXiv#1{\href {http://arxiv.org/abs/#1} {{\tt arXiv:#1}}}
\def\mn@eprint@dblp#1{\href {http://dblp.uni-trier.de/rec/bibtex/#1.xml}
  {dblp:#1}}
\def\mn@eprint@#1:#2:#3:#4\@nil{\def\@tempa {#1}\def\@tempb {#2}\def\@tempc
  {#3}\ifx \@tempc \@empty \let \@tempc \@tempb \let \@tempb \@tempa \fi \ifx
  \@tempb \@empty \def\@tempb {arXiv}\fi \@ifundefined
  {mn@eprint@\@tempb}{\@tempb:\@tempc}{\expandafter \expandafter \csname
  mn@eprint@\@tempb\endcsname \expandafter{\@tempc}}}

\bibitem[\protect\citeauthoryear{{Alecian}}{{Alecian}}{2015}]{alecian:2015}
{Alecian} G.,  2015, \mn@doi [\mnras] {10.1093/mnras/stv2205}, \href
  {http://adsabs.harvard.edu/abs/2015MNRAS.454.3143A} {454, 3143}

\bibitem[\protect\citeauthoryear{{Alecian}, {Tkachenko}, {Neiner}, {Folsom}  \&
  {Leroy}}{{Alecian} et~al.}{2016}]{alecian:2016}
{Alecian} E.,  {Tkachenko} A.,  {Neiner} C.,  {Folsom} C.~P.,   {Leroy} B.,
  2016, \mn@doi [\aap] {10.1051/0004-6361/201527355}, \href
  {http://adsabs.harvard.edu/abs/2016A%26A...589A..47A} {589, A47}

\bibitem[\protect\citeauthoryear{{Amarsi}, {Nordlander}, {Barklem}, {Asplund},
  {Collet}  \& {Lind}}{{Amarsi} et~al.}{2018a}]{amarsi:2018}
{Amarsi} A.~M.,  {Nordlander} T.,  {Barklem} P.~S.,  {Asplund} M.,  {Collet}
  R.,   {Lind} K.,  2018a, \mn@doi [\aap] {10.1051/0004-6361/201732546}, \href
  {https://ui.adsabs.harvard.edu/abs/2018A&A...615A.139A} {615, A139}

\bibitem[\protect\citeauthoryear{{Amarsi}, {Barklem}, {Asplund}, {Collet}  \&
  {Zatsarinny}}{{Amarsi} et~al.}{2018b}]{amarsi:2018a}
{Amarsi} A.~M.,  {Barklem} P.~S.,  {Asplund} M.,  {Collet} R.,   {Zatsarinny}
  O.,  2018b, \mn@doi [\aap] {10.1051/0004-6361/201832770}, \href
  {https://ui.adsabs.harvard.edu/abs/2018A&A...616A..89A} {616, A89}

\bibitem[\protect\citeauthoryear{{Aslan} et~al.,}{{Aslan}
  et~al.}{2001}]{aslan:2001}
{Aslan} Z.,  et~al., 2001, \mn@doi [Astronomy Letters] {10.1134/1.1374679},
  \href {https://ui.adsabs.harvard.edu/abs/2001AstL...27..398A} {27, 398}

\bibitem[\protect\citeauthoryear{{Asplund}, {Amarsi}  \& {Grevesse}}{{Asplund}
  et~al.}{2021}]{asplund:2021}
{Asplund} M.,  {Amarsi} A.~M.,   {Grevesse} N.,  2021, \mn@doi [\aap]
  {10.1051/0004-6361/202140445}, \href
  {https://ui.adsabs.harvard.edu/abs/2021A&A...653A.141A} {653, A141}

\bibitem[\protect\citeauthoryear{{Auri{\`e}re} et~al.,}{{Auri{\`e}re}
  et~al.}{2007}]{auriere:2007}
{Auri{\`e}re} M.,  et~al., 2007, \mn@doi [\aap] {10.1051/0004-6361:20078189},
  \href {http://adsabs.harvard.edu/abs/2007A%26A...475.1053A} {475, 1053}

\bibitem[\protect\citeauthoryear{{Babcock}}{{Babcock}}{1958}]{babcock:1958a}
{Babcock} H.~W.,  1958, \apj, \href
  {http://adsabs.harvard.edu/abs/1958ApJ...128..228B} {128, 228}

\bibitem[\protect\citeauthoryear{{Babel} \& {Michaud}}{{Babel} \&
  {Michaud}}{1991}]{babel:1991}
{Babel} J.,  {Michaud} G.,  1991, \mn@doi [\apj] {10.1086/169591}, \href
  {http://adsabs.harvard.edu/abs/1991ApJ...366..560B} {366, 560}

\bibitem[\protect\citeauthoryear{{Bagnulo}, {Landolfi}, {Landstreet}, {Landi
  Degl'Innocenti}, {Fossati}  \& {Sterzik}}{{Bagnulo}
  et~al.}{2009}]{bagnulo:2009}
{Bagnulo} S.,  {Landolfi} M.,  {Landstreet} J.~D.,  {Landi Degl'Innocenti} E.,
  {Fossati} L.,   {Sterzik} M.,  2009, \mn@doi [\pasp] {10.1086/605654}, \href
  {http://adsabs.harvard.edu/abs/2009PASP..121..993B} {121, 993}

\bibitem[\protect\citeauthoryear{{Bailey}, {Landstreet}  \& {Bagnulo}}{{Bailey}
  et~al.}{2014}]{bailey:2014a}
{Bailey} J.~D.,  {Landstreet} J.~D.,   {Bagnulo} S.,  2014, \mn@doi [\aap]
  {10.1051/0004-6361/201322853}, \href
  {https://ui.adsabs.harvard.edu/abs/2014A&A...561A.147B} {561, A147}

\bibitem[\protect\citeauthoryear{{Bischoff} et~al.,}{{Bischoff}
  et~al.}{2017}]{bischoff:2017}
{Bischoff} R.,  et~al., 2017, \mn@doi [Astronomische Nachrichten]
  {10.1002/asna.201713365}, \href
  {https://ui.adsabs.harvard.edu/abs/2017AN....338..671B} {338, 671}

\bibitem[\protect\citeauthoryear{{Blaz{\`e}re} et~al.,}{{Blaz{\`e}re}
  et~al.}{2016}]{blazere:2016}
{Blaz{\`e}re} A.,  et~al., 2016, \mn@doi [\aap] {10.1051/0004-6361/201527556},
  \href {https://ui.adsabs.harvard.edu/abs/2016A&A...586A..97B} {586, A97}

\bibitem[\protect\citeauthoryear{{Blaz{\`e}re}, {Petit}, {Neiner}, {Folsom},
  {Kochukhov}, {Mathis}, {Deal}  \& {Landstreet}}{{Blaz{\`e}re}
  et~al.}{2020}]{blazere:2020}
{Blaz{\`e}re} A.,  {Petit} P.,  {Neiner} C.,  {Folsom} C.,  {Kochukhov} O.,
  {Mathis} S.,  {Deal} M.,   {Landstreet} J.,  2020, \mn@doi [\mnras]
  {10.1093/mnras/stz3637}, \href
  {https://ui.adsabs.harvard.edu/abs/2020MNRAS.492.5794B} {492, 5794}

\bibitem[\protect\citeauthoryear{{Braithwaite} \& {Spruit}}{{Braithwaite} \&
  {Spruit}}{2004}]{braithwaite:2004}
{Braithwaite} J.,  {Spruit} H.~C.,  2004, \mn@doi [\nat] {10.1038/nature02934},
  \href {http://adsabs.harvard.edu/abs/2004Natur.431..819B} {431, 819}

\bibitem[\protect\citeauthoryear{{Burke} \& {Barr}}{{Burke} \&
  {Barr}}{1981}]{burke:1981}
{Burke} E.~W. J.,  {Barr} T.~H.,  1981, \mn@doi [\pasp] {10.1086/130836}, \href
  {https://ui.adsabs.harvard.edu/abs/1981PASP...93..344B} {93, 344}

\bibitem[\protect\citeauthoryear{{Cantiello} \& {Braithwaite}}{{Cantiello} \&
  {Braithwaite}}{2011}]{cantiello:2011}
{Cantiello} M.,  {Braithwaite} J.,  2011, \mn@doi [\aap]
  {10.1051/0004-6361/201117512}, \href
  {http://adsabs.harvard.edu/abs/2011A%26A...534A.140C} {534, A140}

\bibitem[\protect\citeauthoryear{{Cantiello} \& {Braithwaite}}{{Cantiello} \&
  {Braithwaite}}{2019}]{cantiello:2019}
{Cantiello} M.,  {Braithwaite} J.,  2019, \mn@doi [\apj]
  {10.3847/1538-4357/ab3924}, \href
  {https://ui.adsabs.harvard.edu/abs/2019ApJ...883..106C} {883, 106}

\bibitem[\protect\citeauthoryear{{Cantiello} et~al.,}{{Cantiello}
  et~al.}{2009}]{cantiello:2009}
{Cantiello} M.,  et~al., 2009, \mn@doi [\aap] {10.1051/0004-6361/200911643},
  \href {http://esoads.eso.org/abs/2009A%26A...499..279C} {499, 279}

\bibitem[\protect\citeauthoryear{{Carrier}, {North}, {Udry}  \&
  {Babel}}{{Carrier} et~al.}{2002}]{carrier:2002}
{Carrier} F.,  {North} P.,  {Udry} S.,   {Babel} J.,  2002, \mn@doi [\aap]
  {10.1051/0004-6361:20021122}, \href
  {http://adsabs.harvard.edu/abs/2002A%26A...394..151C} {394, 151}

\bibitem[\protect\citeauthoryear{{Chini}, {Hoffmeister}, {Nasseri}, {Stahl}  \&
  {Zinnecker}}{{Chini} et~al.}{2012}]{chini:2012}
{Chini} R.,  {Hoffmeister} V.~H.,  {Nasseri} A.,  {Stahl} O.,   {Zinnecker} H.,
   2012, \mn@doi [\mnras] {10.1111/j.1365-2966.2012.21317.x}, \href
  {https://ui.adsabs.harvard.edu/abs/2012MNRAS.424.1925C} {424, 1925}

\bibitem[\protect\citeauthoryear{{Choi}, {Dotter}, {Conroy}, {Cantiello},
  {Paxton}  \& {Johnson}}{{Choi} et~al.}{2016}]{choi:2016}
{Choi} J.,  {Dotter} A.,  {Conroy} C.,  {Cantiello} M.,  {Paxton} B.,
  {Johnson} B.~D.,  2016, \mn@doi [\apj] {10.3847/0004-637X/823/2/102}, \href
  {https://ui.adsabs.harvard.edu/abs/2016ApJ...823..102C} {823, 102}

\bibitem[\protect\citeauthoryear{{Cohen}, {Wheaton}  \& {Megeath}}{{Cohen}
  et~al.}{2003}]{cohen:2003}
{Cohen} M.,  {Wheaton} W.~A.,   {Megeath} S.~T.,  2003, \mn@doi [\aj]
  {10.1086/376474}, \href
  {https://ui.adsabs.harvard.edu/abs/2003AJ....126.1090C} {126, 1090}

\bibitem[\protect\citeauthoryear{{Cutri} et~al.,}{{Cutri}
  et~al.}{2003}]{cutri:2003}
{Cutri} R.~M.,  et~al., 2003, VizieR Online Data Catalog, \href
  {https://ui.adsabs.harvard.edu/abs/2003yCat.2246....0C} {p. II/246}

\bibitem[\protect\citeauthoryear{{Deutsch}}{{Deutsch}}{1947}]{deutsch:1947}
{Deutsch} A.~J.,  1947, \mn@doi [\apj] {10.1086/144904}, \href
  {http://cdsads.u-strasbg.fr/abs/1947ApJ...105..283D} {105, 283}

\bibitem[\protect\citeauthoryear{{Donati} \& {Landstreet}}{{Donati} \&
  {Landstreet}}{2009}]{donati:2009}
{Donati} J.-F.,  {Landstreet} J.~D.,  2009, \mn@doi [\araa]
  {10.1146/annurev-astro-082708-101833}, \href
  {http://adsabs.harvard.edu/abs/2009ARA%26A..47..333D} {47, 333}

\bibitem[\protect\citeauthoryear{{Donati}, {Semel}  \& {Rees}}{{Donati}
  et~al.}{1992}]{donati:1992}
{Donati} J.-F.,  {Semel} M.,   {Rees} D.~E.,  1992, \aap, \href
  {http://adsabs.harvard.edu/abs/1992A%26A...265..669D} {265, 669}

\bibitem[\protect\citeauthoryear{{Donati}, {Semel}, {Carter}, {Rees}  \&
  {Collier Cameron}}{{Donati} et~al.}{1997}]{donati:1997}
{Donati} J.~F.,  {Semel} M.,  {Carter} B.~D.,  {Rees} D.~E.,   {Collier
  Cameron} A.,  1997, \mn@doi [\mnras] {10.1093/mnras/291.4.658}, \href
  {https://ui.adsabs.harvard.edu/abs/1997MNRAS.291..658D} {291, 658}

\bibitem[\protect\citeauthoryear{{Donati}, {Collier Cameron}, {Hussain}  \&
  {Semel}}{{Donati} et~al.}{1999}]{donati:1999}
{Donati} J.-F.,  {Collier Cameron} A.,  {Hussain} G.~A.~J.,   {Semel} M.,
  1999, \mnras, \href {http://adsabs.harvard.edu/abs/1999MNRAS.302..437D} {302,
  437}

\bibitem[\protect\citeauthoryear{{Dotter}}{{Dotter}}{2016}]{dotter:2016}
{Dotter} A.,  2016, \mn@doi [\apjs] {10.3847/0067-0049/222/1/8}, \href
  {https://ui.adsabs.harvard.edu/abs/2016ApJS..222....8D} {222, 8}

\bibitem[\protect\citeauthoryear{{Farrell}, {Jermyn}, {Cantiello}  \&
  {Foreman-Mackey}}{{Farrell} et~al.}{2022}]{farrell:2022}
{Farrell} E.,  {Jermyn} A.~S.,  {Cantiello} M.,   {Foreman-Mackey} D.,  2022,
  \mn@doi [\apj] {10.3847/1538-4357/ac8423}, \href
  {https://ui.adsabs.harvard.edu/abs/2022ApJ...938...10F} {938, 10}

\bibitem[\protect\citeauthoryear{{Gaia Collaboration} et~al.,}{{Gaia
  Collaboration} et~al.}{2021}]{gaia-collaboration:2021}
{Gaia Collaboration} et~al., 2021, \mn@doi [\aap]
  {10.1051/0004-6361/202039657}, \href
  {https://ui.adsabs.harvard.edu/abs/2021A&A...649A...1G} {649, A1}

\bibitem[\protect\citeauthoryear{{Glagolevskij}}{{Glagolevskij}}{2019}]{glagolevskij:2019}
{Glagolevskij} Y.~V.,  2019, \mn@doi [Astrophysical Bulletin]
  {10.1134/S1990341319010073}, \href
  {https://ui.adsabs.harvard.edu/abs/2019AstBu..74...66G} {74, 66}

\bibitem[\protect\citeauthoryear{{Groote} \& {Kaufmann}}{{Groote} \&
  {Kaufmann}}{1983}]{groote:1983}
{Groote} D.,  {Kaufmann} J.~P.,  1983, \aaps, \href
  {http://adsabs.harvard.edu/abs/1983A%26AS...53...91G} {53, 91}

\bibitem[\protect\citeauthoryear{{Hatzes}}{{Hatzes}}{1991}]{hatzes:1991a}
{Hatzes} A.~P.,  1991, \mn@doi [\mnras] {10.1093/mnras/253.1.89}, \href
  {http://adsabs.harvard.edu/abs/1991MNRAS.253...89H} {253, 89}

\bibitem[\protect\citeauthoryear{{Hubrig}, {North}, {Sch{\"o}ller}  \&
  {Mathys}}{{Hubrig} et~al.}{2006}]{hubrig:2006b}
{Hubrig} S.,  {North} P.,  {Sch{\"o}ller} M.,   {Mathys} G.,  2006, \mn@doi
  [Astronomische Nachrichten] {10.1002/asna.200610535}, \href
  {http://adsabs.harvard.edu/abs/2006AN....327..289H} {327, 289}

\bibitem[\protect\citeauthoryear{{Jermyn} \& {Cantiello}}{{Jermyn} \&
  {Cantiello}}{2020}]{jermyn:2020}
{Jermyn} A.~S.,  {Cantiello} M.,  2020, \mn@doi [\apj]
  {10.3847/1538-4357/ab9e70}, \href
  {https://ui.adsabs.harvard.edu/abs/2020ApJ...900..113J} {900, 113}

\bibitem[\protect\citeauthoryear{{Kochukhov}}{{Kochukhov}}{2007}]{kochukhov:2007d}
{Kochukhov} O.,  2007, in {Romanyuk} I.~I.,  {Kudryavtsev} D.~O.,  eds, Physics
  of Magnetic Stars. pp 109--118

\bibitem[\protect\citeauthoryear{{Kochukhov}}{{Kochukhov}}{2017}]{kochukhov:2017}
{Kochukhov} O.,  2017, \mn@doi [\aap] {10.1051/0004-6361/201629768}, \href
  {http://adsabs.harvard.edu/abs/2017A%26A...597A..58K} {597, A58}

\bibitem[\protect\citeauthoryear{{Kochukhov}}{{Kochukhov}}{2018}]{kochukhov:2018}
{Kochukhov} O.,  2018, {BinMag: Widget for comparing stellar observed with
  theoretical spectra}, Astrophysics Source Code Library, record ascl:1805.015
  (\mn@eprint {ascl} {1805.015})

\bibitem[\protect\citeauthoryear{{Kochukhov} \& {Bagnulo}}{{Kochukhov} \&
  {Bagnulo}}{2006}]{kochukhov:2006}
{Kochukhov} O.,  {Bagnulo} S.,  2006, \mn@doi [\aap]
  {10.1051/0004-6361:20054596}, \href
  {http://adsabs.harvard.edu/abs/2006A%26A...450..763K} {450, 763}

\bibitem[\protect\citeauthoryear{{Kochukhov} \& {Lavail}}{{Kochukhov} \&
  {Lavail}}{2017}]{kochukhov:2017c}
{Kochukhov} O.,  {Lavail} A.,  2017, \mn@doi [\apjl]
  {10.3847/2041-8213/835/1/L4}, \href
  {https://ui.adsabs.harvard.edu/abs/2017ApJ...835L...4K} {835, L4}

\bibitem[\protect\citeauthoryear{{Kochukhov} \& {Piskunov}}{{Kochukhov} \&
  {Piskunov}}{2002}]{kochukhov:2002c}
{Kochukhov} O.,  {Piskunov} N.,  2002, \mn@doi [\aap]
  {10.1051/0004-6361:20020300}, \href
  {https://ui.adsabs.harvard.edu/abs/2002A&A...388..868K} {388, 868}

\bibitem[\protect\citeauthoryear{{Kochukhov} \& {Reiners}}{{Kochukhov} \&
  {Reiners}}{2020}]{kochukhov:2020a}
{Kochukhov} O.,  {Reiners} A.,  2020, \mn@doi [\apj]
  {10.3847/1538-4357/abb2a2}, \href
  {https://ui.adsabs.harvard.edu/abs/2020ApJ...902...43K} {902, 43}

\bibitem[\protect\citeauthoryear{{Kochukhov} \& {Wade}}{{Kochukhov} \&
  {Wade}}{2010}]{kochukhov:2010}
{Kochukhov} O.,  {Wade} G.~A.,  2010, \mn@doi [\aap]
  {10.1051/0004-6361/200913860}, \href
  {https://ui.adsabs.harvard.edu/abs/2010A&A...513A..13K} {513, A13}

\bibitem[\protect\citeauthoryear{{Kochukhov}, {Drake}, {Piskunov}  \& {de la
  Reza}}{{Kochukhov} et~al.}{2004}]{kochukhov:2004e}
{Kochukhov} O.,  {Drake} N.~A.,  {Piskunov} N.,   {de la Reza} R.,  2004,
  \mn@doi [\aap] {10.1051/0004-6361:20040517}, \href
  {http://adsabs.harvard.edu/abs/2004A%26A...424..935K} {424, 935}

\bibitem[\protect\citeauthoryear{{Kochukhov}, {Tsymbal}, {Ryabchikova},
  {Makaganyk}  \& {Bagnulo}}{{Kochukhov} et~al.}{2006}]{kochukhov:2006b}
{Kochukhov} O.,  {Tsymbal} V.,  {Ryabchikova} T.,  {Makaganyk} V.,   {Bagnulo}
  S.,  2006, \mn@doi [\aap] {10.1051/0004-6361:20065607}, \href
  {http://adsabs.harvard.edu/abs/2006A%26A...460..831K} {460, 831}

\bibitem[\protect\citeauthoryear{{Kochukhov}, {Adelman}, {Gulliver}  \&
  {Piskunov}}{{Kochukhov} et~al.}{2007}]{kochukhov:2007b}
{Kochukhov} O.,  {Adelman} S.~J.,  {Gulliver} A.~F.,   {Piskunov} N.,  2007,
  \mn@doi [Nature Physics] {10.1038/nphys648}, \href
  {http://esoads.eso.org/abs/2007NatPh...3..526K} {3, 526}

\bibitem[\protect\citeauthoryear{{Kochukhov}, {Makaganiuk}  \&
  {Piskunov}}{{Kochukhov} et~al.}{2010}]{kochukhov:2010a}
{Kochukhov} O.,  {Makaganiuk} V.,   {Piskunov} N.,  2010, \mn@doi [\aap]
  {10.1051/0004-6361/201015429}, \href
  {https://ui.adsabs.harvard.edu/abs/2010A&A...524A...5K} {524, A5}

\bibitem[\protect\citeauthoryear{{Kochukhov} et~al.,}{{Kochukhov}
  et~al.}{2011}]{kochukhov:2011b}
{Kochukhov} O.,  et~al., 2011, \mn@doi [\aap] {10.1051/0004-6361/201117970},
  \href {http://cdsads.u-strasbg.fr/abs/2011A%26A...534L..13K} {534, L13}

\bibitem[\protect\citeauthoryear{{Kochukhov}, {L{\"u}ftinger}, {Neiner},
  {Alecian}  \& {MiMeS Collaboration}}{{Kochukhov}
  et~al.}{2014}]{kochukhov:2014}
{Kochukhov} O.,  {L{\"u}ftinger} T.,  {Neiner} C.,  {Alecian} E.,   {MiMeS
  Collaboration} 2014, \mn@doi [\aap] {10.1051/0004-6361/201423472}, \href
  {http://adsabs.harvard.edu/abs/2014A%26A...565A..83K} {565, A83}

\bibitem[\protect\citeauthoryear{{Kochukhov}, {Silvester}, {Bailey},
  {Landstreet}  \& {Wade}}{{Kochukhov} et~al.}{2017}]{kochukhov:2017a}
{Kochukhov} O.,  {Silvester} J.,  {Bailey} J.~D.,  {Landstreet} J.~D.,   {Wade}
  G.~A.,  2017, \mn@doi [\aap] {10.1051/0004-6361/201730919}, \href
  {http://adsabs.harvard.edu/abs/2017A%26A...605A..13K} {605, A13}

\bibitem[\protect\citeauthoryear{{Kochukhov}, {Shultz}  \&
  {Neiner}}{{Kochukhov} et~al.}{2019}]{kochukhov:2019}
{Kochukhov} O.,  {Shultz} M.,   {Neiner} C.,  2019, \mn@doi [\aap]
  {10.1051/0004-6361/201834279}, \href
  {http://adsabs.harvard.edu/abs/2019A%26A...621A..47K} {621, A47}

\bibitem[\protect\citeauthoryear{{Kochukhov}, {Khalack}, {Kobzar}, {Neiner},
  {Paunzen}, {Labadie-Bartz}  \& {David-Uraz}}{{Kochukhov}
  et~al.}{2021}]{kochukhov:2021b}
{Kochukhov} O.,  {Khalack} V.,  {Kobzar} O.,  {Neiner} C.,  {Paunzen} E.,
  {Labadie-Bartz} J.,   {David-Uraz} A.,  2021, \mn@doi [\mnras]
  {10.1093/mnras/stab2107}, \href
  {https://ui.adsabs.harvard.edu/abs/2021MNRAS.506.5328K} {506, 5328}

\bibitem[\protect\citeauthoryear{{Kochukhov}, {Papakonstantinou}  \&
  {Neiner}}{{Kochukhov} et~al.}{2022}]{kochukhov:2022}
{Kochukhov} O.,  {Papakonstantinou} N.,   {Neiner} C.,  2022, \mn@doi [\mnras]
  {10.1093/mnras/stac066}, \href
  {https://ui.adsabs.harvard.edu/abs/2022MNRAS.510.5821K} {510, 5821}

\bibitem[\protect\citeauthoryear{{Korhonen} et~al.,}{{Korhonen}
  et~al.}{2013}]{korhonen:2013}
{Korhonen} H.,  et~al., 2013, \mn@doi [\aap] {10.1051/0004-6361/201220951},
  \href {http://adsabs.harvard.edu/abs/2013A%26A...553A..27K} {553, A27}

\bibitem[\protect\citeauthoryear{{Kurucz}}{{Kurucz}}{1995}]{kurucz:1995}
{Kurucz} R.~L.,  1995, in {Adelman} S.~J.,  {Wiese} W.~L.,  eds,  Astronomical
  Society of the Pacific Conference Series Vol. 78, Astrophysical Applications
  of Powerful New Databases. p.~205

\bibitem[\protect\citeauthoryear{{Lallement}, {Babusiaux}, {Vergely}, {Katz},
  {Arenou}, {Valette}, {Hottier}  \& {Capitanio}}{{Lallement}
  et~al.}{2019}]{lallement:2019}
{Lallement} R.,  {Babusiaux} C.,  {Vergely} J.~L.,  {Katz} D.,  {Arenou} F.,
  {Valette} B.,  {Hottier} C.,   {Capitanio} L.,  2019, \mn@doi [\aap]
  {10.1051/0004-6361/201834695}, \href
  {https://ui.adsabs.harvard.edu/abs/2019A&A...625A.135L} {625, A135}

\bibitem[\protect\citeauthoryear{{Landi Degl'Innocenti} \& {Landolfi}}{{Landi
  Degl'Innocenti} \& {Landolfi}}{2004}]{polarization:2004}
{Landi Degl'Innocenti} E.,  {Landolfi} M.,  2004, {Polarization in Spectral
  Lines}.
 Astrophysics and Space Science Library Vol. 307, Kluwer Academic Publishers

\bibitem[\protect\citeauthoryear{{Leenaarts} \& {Carlsson}}{{Leenaarts} \&
  {Carlsson}}{2009}]{leenaarts:2009}
{Leenaarts} J.,  {Carlsson} M.,  2009, in {Lites} B.,  {Cheung} M.,  {Magara}
  T.,  {Mariska} J.,   {Reeves} K.,  eds,  Astronomical Society of the Pacific
  Conference Series Vol. 415, The Second Hinode Science Meeting: Beyond
  Discovery-Toward Understanding. p.~87

\bibitem[\protect\citeauthoryear{{Lehtinen} et~al.,}{{Lehtinen}
  et~al.}{2022}]{lehtinen:2022}
{Lehtinen} J.~J.,  et~al., 2022, \mn@doi [\aap] {10.1051/0004-6361/201936780},
  \href {https://ui.adsabs.harvard.edu/abs/2022A&A...660A.141L} {660, A141}

\bibitem[\protect\citeauthoryear{{L{\"u}ftinger} et~al.,}{{L{\"u}ftinger}
  et~al.}{2010}]{luftinger:2010a}
{L{\"u}ftinger} T.,  et~al., 2010, \mn@doi [\aap]
  {10.1051/0004-6361/200912239}, \href
  {http://adsabs.harvard.edu/abs/2010A%26A...509A..43L} {509, A43}

\bibitem[\protect\citeauthoryear{{MacDonald} \& {Petit}}{{MacDonald} \&
  {Petit}}{2019}]{macdonald:2019}
{MacDonald} J.,  {Petit} V.,  2019, \mn@doi [\mnras] {10.1093/mnras/stz1545},
  \href {https://ui.adsabs.harvard.edu/abs/2019MNRAS.487.3904M} {487, 3904}

\bibitem[\protect\citeauthoryear{{Markwardt}}{{Markwardt}}{2009}]{markwardt:2009}
{Markwardt} C.~B.,  2009, in {Bohlender} D.~A.,  {Durand} D.,   {Dowler} P.,
  eds,  Astronomical Society of the Pacific Conference Series Vol. 411,
  Astronomical Data Analysis Software and Systems XVIII. p.~251 (\mn@eprint
  {arXiv} {0902.2850})

\bibitem[\protect\citeauthoryear{{Mason}, {Wycoff}, {Hartkopf}, {Douglass}  \&
  {Worley}}{{Mason} et~al.}{2001}]{mason:2001}
{Mason} B.~D.,  {Wycoff} G.~L.,  {Hartkopf} W.~I.,  {Douglass} G.~G.,
  {Worley} C.~E.,  2001, \mn@doi [\aj] {10.1086/323920}, \href
  {https://ui.adsabs.harvard.edu/abs/2001AJ....122.3466M} {122, 3466}

\bibitem[\protect\citeauthoryear{{Mathys}}{{Mathys}}{1991}]{mathys:1991}
{Mathys} G.,  1991, \aaps, \href
  {http://adsabs.harvard.edu/abs/1991A%26AS...89..121M} {89, 121}

\bibitem[\protect\citeauthoryear{{Mestel}}{{Mestel}}{2003}]{mestel:2003}
{Mestel} L.,  2003, in {L.~A.~Balona, H.~F.~Henrichs, \& R.~Medupe} ed.,  Vol.
  305, Astronomical Society of the Pacific Conference Series. pp 3--15

\bibitem[\protect\citeauthoryear{{Michaud}, {Charland}  \&
  {Megessier}}{{Michaud} et~al.}{1981}]{michaud:1981}
{Michaud} G.,  {Charland} Y.,   {Megessier} C.,  1981, \aap, \href
  {http://adsabs.harvard.edu/abs/1981A%26A...103..244M} {103, 244}

\bibitem[\protect\citeauthoryear{{Montegriffo} et~al.,}{{Montegriffo}
  et~al.}{2022}]{montegriffo:2022}
{Montegriffo} P.,  et~al., 2022, arXiv e-prints, \href
  {https://ui.adsabs.harvard.edu/abs/2022arXiv220606205M} {p. arXiv:2206.06205}

\bibitem[\protect\citeauthoryear{{Morin} et~al.,}{{Morin}
  et~al.}{2008}]{morin:2008}
{Morin} J.,  et~al., 2008, \mn@doi [\mnras] {10.1111/j.1365-2966.2008.13809.x},
  \href {https://ui.adsabs.harvard.edu/abs/2008MNRAS.390..567M} {390, 567}

\bibitem[\protect\citeauthoryear{{Moss}}{{Moss}}{2004}]{moss:2004}
{Moss} D.,  2004, in {Zverko} J.,  {Ziznovsky} J.,  {Adelman} S.~J.,   {Weiss}
  W.~W.,  eds,  IAU Symposium Vol. 224, The A-Star Puzzle. pp 245--252

\bibitem[\protect\citeauthoryear{{Neiner}, {Mathis}, {Alecian}, {Emeriau},
  {Grunhut}, {BinaMIcS}  \& {MiMeS Collaborations}}{{Neiner}
  et~al.}{2015}]{neiner:2015}
{Neiner} C.,  {Mathis} S.,  {Alecian} E.,  {Emeriau} C.,  {Grunhut} J.,
  {BinaMIcS}  {MiMeS Collaborations} 2015, in {Nagendra} K.~N.,  {Bagnulo} S.,
  {Centeno} R.,   {Jes{\'u}s Mart{\'{\i}}nez Gonz{\'a}lez} M.,  eds,  IAU
  Symposium Vol. 305, Polarimetry. pp 61--66

\bibitem[\protect\citeauthoryear{{Netopil}, {Paunzen}, {Maitzen}, {North}  \&
  {Hubrig}}{{Netopil} et~al.}{2008}]{netopil:2008}
{Netopil} M.,  {Paunzen} E.,  {Maitzen} H.~M.,  {North} P.,   {Hubrig} S.,
  2008, \mn@doi [\aap] {10.1051/0004-6361:200810325}, \href
  {https://ui.adsabs.harvard.edu/abs/2008A&A...491..545N} {491, 545}

\bibitem[\protect\citeauthoryear{{Netopil}, {Paunzen}, {H{\"u}mmerich}  \&
  {Bernhard}}{{Netopil} et~al.}{2017}]{netopil:2017}
{Netopil} M.,  {Paunzen} E.,  {H{\"u}mmerich} S.,   {Bernhard} K.,  2017,
  \mn@doi [\mnras] {10.1093/mnras/stx674}, \href
  {https://ui.adsabs.harvard.edu/abs/2017MNRAS.468.2745N} {468, 2745}

\bibitem[\protect\citeauthoryear{{Paunzen}, {Sup{\'\i}kov{\'a}}, {Bernhard},
  {H{\"u}mmerich}  \& {Pri{\v{s}}egen}}{{Paunzen} et~al.}{2021}]{paunzen:2021b}
{Paunzen} E.,  {Sup{\'\i}kov{\'a}} J.,  {Bernhard} K.,  {H{\"u}mmerich} S.,
  {Pri{\v{s}}egen} M.,  2021, \mn@doi [\mnras] {10.1093/mnras/stab1100}, \href
  {https://ui.adsabs.harvard.edu/abs/2021MNRAS.504.3758P} {504, 3758}

\bibitem[\protect\citeauthoryear{{Paxton}, {Bildsten}, {Dotter}, {Herwig},
  {Lesaffre}  \& {Timmes}}{{Paxton} et~al.}{2011}]{paxton:2011}
{Paxton} B.,  {Bildsten} L.,  {Dotter} A.,  {Herwig} F.,  {Lesaffre} P.,
  {Timmes} F.,  2011, \mn@doi [\apjs] {10.1088/0067-0049/192/1/3}, \href
  {https://ui.adsabs.harvard.edu/abs/2011ApJS..192....3P} {192, 3}

\bibitem[\protect\citeauthoryear{{Petit}, {Donati}  \& {Collier
  Cameron}}{{Petit} et~al.}{2002}]{petit:2002}
{Petit} P.,  {Donati} J.~F.,   {Collier Cameron} A.,  2002, \mn@doi [\mnras]
  {10.1046/j.1365-8711.2002.05529.x}, \href
  {https://ui.adsabs.harvard.edu/abs/2002MNRAS.334..374P} {334, 374}

\bibitem[\protect\citeauthoryear{{Petit} et~al.,}{{Petit}
  et~al.}{2010}]{petit:2010}
{Petit} P.,  et~al., 2010, \mn@doi [\aap] {10.1051/0004-6361/201015307}, \href
  {http://cdsads.u-strasbg.fr/abs/2010A%26A...523A..41P} {523, A41}

\bibitem[\protect\citeauthoryear{{Petit} et~al.,}{{Petit}
  et~al.}{2011}]{petit:2011}
{Petit} P.,  et~al., 2011, \mn@doi [\aap] {10.1051/0004-6361/201117573}, \href
  {http://adsabs.harvard.edu/abs/2011A%26A...532L..13P} {532, L13}

\bibitem[\protect\citeauthoryear{{Petit}, {Louge}, {Th{\'e}ado}, {Paletou},
  {Manset}, {Morin}, {Marsden}  \& {Jeffers}}{{Petit}
  et~al.}{2014}]{petit:2014}
{Petit} P.,  {Louge} T.,  {Th{\'e}ado} S.,  {Paletou} F.,  {Manset} N.,
  {Morin} J.,  {Marsden} S.~C.,   {Jeffers} S.~V.,  2014, \mn@doi [\pasp]
  {10.1086/676976}, \href {http://adsabs.harvard.edu/abs/2014PASP..126..469P}
  {126, 469}

\bibitem[\protect\citeauthoryear{{Piskunov} \& {Kochukhov}}{{Piskunov} \&
  {Kochukhov}}{2002}]{piskunov:2002a}
{Piskunov} N.,  {Kochukhov} O.,  2002, \mn@doi [\aap]
  {10.1051/0004-6361:20011517}, \href
  {https://ui.adsabs.harvard.edu/abs/2002A&A...381..736P} {381, 736}

\bibitem[\protect\citeauthoryear{{Piskunov} \& {Valenti}}{{Piskunov} \&
  {Valenti}}{2017}]{piskunov:2017}
{Piskunov} N.,  {Valenti} J.~A.,  2017, \mn@doi [\aap]
  {10.1051/0004-6361/201629124}, \href
  {http://adsabs.harvard.edu/abs/2017A%26A...597A..16P} {597, A16}

\bibitem[\protect\citeauthoryear{{Preston}}{{Preston}}{1974}]{preston:1974}
{Preston} G.~W.,  1974, \mn@doi [\araa] {10.1146/annurev.aa.12.090174.001353},
  \href {http://adsabs.harvard.edu/abs/1974ARA%26A..12..257P} {12, 257}

\bibitem[\protect\citeauthoryear{{Pr{\v{s}}a} et~al.,}{{Pr{\v{s}}a}
  et~al.}{2016}]{prsa:2016a}
{Pr{\v{s}}a} A.,  et~al., 2016, \mn@doi [\aj] {10.3847/0004-6256/152/2/41},
  \href {https://ui.adsabs.harvard.edu/abs/2016AJ....152...41P} {152, 41}

\bibitem[\protect\citeauthoryear{{Renson} \& {Manfroid}}{{Renson} \&
  {Manfroid}}{2009}]{renson:2009}
{Renson} P.,  {Manfroid} J.,  2009, \mn@doi [\aap]
  {10.1051/0004-6361/200810788}, \href
  {http://adsabs.harvard.edu/abs/2009A%26A...498..961R} {498, 961}

\bibitem[\protect\citeauthoryear{{Rice}, {Wehlau}  \& {Holmgren}}{{Rice}
  et~al.}{1997}]{rice:1997}
{Rice} J.~B.,  {Wehlau} W.~H.,   {Holmgren} D.~E.,  1997, \aap, \href
  {http://adsabs.harvard.edu/abs/1997A%26A...326..988R} {326, 988}

\bibitem[\protect\citeauthoryear{{Ros{\'e}n}, {Kochukhov}, {Alecian}, {Neiner},
  {Morin}, {Wade}  \& {BinaMIcS Collaboration}}{{Ros{\'e}n}
  et~al.}{2018}]{rosen:2018}
{Ros{\'e}n} L.,  {Kochukhov} O.,  {Alecian} E.,  {Neiner} C.,  {Morin} J.,
  {Wade} G.~A.,   {BinaMIcS Collaboration} 2018, \mn@doi [\aap]
  {10.1051/0004-6361/201731706}, \href
  {http://adsabs.harvard.edu/abs/2018A%26A...613A..60R} {613, A60}

\bibitem[\protect\citeauthoryear{{Rufener}}{{Rufener}}{1988}]{rufener:1988}
{Rufener} F.,  1988, {Catalogue of stars measured in the Geneva Observatory
  photometric system : 4 : 1988}.
Sauverny: Observatoire de Geneve, 1988

\bibitem[\protect\citeauthoryear{{Rufener} \& {Nicolet}}{{Rufener} \&
  {Nicolet}}{1988}]{rufener:1988a}
{Rufener} F.,  {Nicolet} B.,  1988, \aap, \href
  {https://ui.adsabs.harvard.edu/abs/1988A&A...206..357R} {206, 357}

\bibitem[\protect\citeauthoryear{{Rusomarov}, {Kochukhov}, {Ryabchikova}  \&
  {Ilyin}}{{Rusomarov} et~al.}{2016}]{rusomarov:2016}
{Rusomarov} N.,  {Kochukhov} O.,  {Ryabchikova} T.,   {Ilyin} I.,  2016,
  \mn@doi [\aap] {10.1051/0004-6361/201527719}, \href
  {http://adsabs.harvard.edu/abs/2016A%26A...588A.138R} {588, A138}

\bibitem[\protect\citeauthoryear{{Rusomarov}, {Kochukhov}  \&
  {Lundin}}{{Rusomarov} et~al.}{2018}]{rusomarov:2018}
{Rusomarov} N.,  {Kochukhov} O.,   {Lundin} A.,  2018, \mn@doi [\aap]
  {10.1051/0004-6361/201731914}, \href
  {http://adsabs.harvard.edu/abs/2018A%26A...609A..88R} {609, A88}

\bibitem[\protect\citeauthoryear{{Ryabchikova}}{{Ryabchikova}}{2005}]{ryabchikova:2005b}
{Ryabchikova} T.~A.,  2005, \mn@doi [Astronomy Letters] {10.1134/1.1940111},
  \href {http://adsabs.harvard.edu/abs/2005AstL...31..388R} {31, 388}

\bibitem[\protect\citeauthoryear{{Ryabchikova}, {Leone}  \&
  {Kochukhov}}{{Ryabchikova} et~al.}{2005}]{ryabchikova:2005a}
{Ryabchikova} T.,  {Leone} F.,   {Kochukhov} O.,  2005, \mn@doi [\aap]
  {10.1051/0004-6361:20041996}, \href
  {http://adsabs.harvard.edu/abs/2005A%26A...438..973R} {438, 973}

\bibitem[\protect\citeauthoryear{{Ryabchikova}, {Piskunov}, {Kurucz},
  {Stempels}, {Heiter}, {Pakhomov}  \& {Barklem}}{{Ryabchikova}
  et~al.}{2015}]{ryabchikova:2015}
{Ryabchikova} T.,  {Piskunov} N.,  {Kurucz} R.~L.,  {Stempels} H.~C.,  {Heiter}
  U.,  {Pakhomov} Y.,   {Barklem} P.~S.,  2015, \mn@doi [\physscr]
  {10.1088/0031-8949/90/5/054005}, \href
  {http://adsabs.harvard.edu/abs/2015PhyS...90e4005R} {90, 054005}

\bibitem[\protect\citeauthoryear{{Schneider}, {Ohlmann}, {Podsiadlowski},
  {R{\"o}pke}, {Balbus}, {Pakmor}  \& {Springel}}{{Schneider}
  et~al.}{2019}]{schneider:2019}
{Schneider} F. R.~N.,  {Ohlmann} S.~T.,  {Podsiadlowski} P.,  {R{\"o}pke}
  F.~K.,  {Balbus} S.~A.,  {Pakmor} R.,   {Springel} V.,  2019, \mn@doi [\nat]
  {10.1038/s41586-019-1621-5}, \href
  {https://ui.adsabs.harvard.edu/abs/2019Natur.574..211S} {574, 211}

\bibitem[\protect\citeauthoryear{{Shulyak}, {Tsymbal}, {Ryabchikova},
  {St{\"u}tz}  \& {Weiss}}{{Shulyak} et~al.}{2004}]{shulyak:2004}
{Shulyak} D.,  {Tsymbal} V.,  {Ryabchikova} T.,  {St{\"u}tz} C.,   {Weiss}
  W.~W.,  2004, \mn@doi [\aap] {10.1051/0004-6361:20034169}, \href
  {http://adsabs.harvard.edu/abs/2004A%26A...428..993S} {428, 993}

\bibitem[\protect\citeauthoryear{{Sikora}, {Wade}, {Power}  \&
  {Neiner}}{{Sikora} et~al.}{2019a}]{sikora:2019}
{Sikora} J.,  {Wade} G.~A.,  {Power} J.,   {Neiner} C.,  2019a, \mn@doi
  [\mnras] {10.1093/mnras/sty3105}, \href
  {https://ui.adsabs.harvard.edu/abs/2019MNRAS.483.2300S} {483, 2300}

\bibitem[\protect\citeauthoryear{{Sikora}, {Wade}, {Power}  \&
  {Neiner}}{{Sikora} et~al.}{2019b}]{sikora:2019a}
{Sikora} J.,  {Wade} G.~A.,  {Power} J.,   {Neiner} C.,  2019b, \mn@doi
  [\mnras] {10.1093/mnras/sty2895}, \href
  {https://ui.adsabs.harvard.edu/abs/2019MNRAS.483.3127S} {483, 3127}

\bibitem[\protect\citeauthoryear{{Silvester}, {Kochukhov}  \&
  {Wade}}{{Silvester} et~al.}{2014a}]{silvester:2014}
{Silvester} J.,  {Kochukhov} O.,   {Wade} G.~A.,  2014a, \mn@doi [\mnras]
  {10.1093/mnras/stu306}, \href
  {http://adsabs.harvard.edu/abs/2014MNRAS.440..182S} {440, 182}

\bibitem[\protect\citeauthoryear{{Silvester}, {Kochukhov}  \&
  {Wade}}{{Silvester} et~al.}{2014b}]{silvester:2014a}
{Silvester} J.,  {Kochukhov} O.,   {Wade} G.~A.,  2014b, \mn@doi [\mnras]
  {10.1093/mnras/stu1531}, \href
  {http://adsabs.harvard.edu/abs/2014MNRAS.444.1442S} {444, 1442}

\bibitem[\protect\citeauthoryear{{Silvester}, {Kochukhov}  \&
  {Wade}}{{Silvester} et~al.}{2015}]{silvester:2015}
{Silvester} J.,  {Kochukhov} O.,   {Wade} G.~A.,  2015, \mn@doi [\mnras]
  {10.1093/mnras/stv1775}, \href
  {http://adsabs.harvard.edu/abs/2015MNRAS.453.2163S} {453, 2163}

\bibitem[\protect\citeauthoryear{{Sitnova}, {Mashonkina}  \&
  {Ryabchikova}}{{Sitnova} et~al.}{2013}]{sitnova:2013}
{Sitnova} T.~M.,  {Mashonkina} L.~I.,   {Ryabchikova} T.~A.,  2013, \mn@doi
  [Astronomy Letters] {10.1134/S1063773713020084}, \href
  {https://ui.adsabs.harvard.edu/abs/2013AstL...39..126S} {39, 126}

\bibitem[\protect\citeauthoryear{{Spruit}}{{Spruit}}{1999}]{spruit:1999}
{Spruit} H.~C.,  1999, \aap, \href
  {https://ui.adsabs.harvard.edu/abs/1999A&A...349..189S} {349, 189}

\bibitem[\protect\citeauthoryear{{Takeda} \& {Honda}}{{Takeda} \&
  {Honda}}{2016}]{takeda:2016}
{Takeda} Y.,  {Honda} S.,  2016, \mn@doi [\pasj] {10.1093/pasj/psw021}, \href
  {https://ui.adsabs.harvard.edu/abs/2016PASJ...68...32T} {68, 32}

\bibitem[\protect\citeauthoryear{{Takeda} \& {Sadakane}}{{Takeda} \&
  {Sadakane}}{1997}]{takeda:1997a}
{Takeda} Y.,  {Sadakane} K.,  1997, \mn@doi [\pasj] {10.1093/pasj/49.3.367},
  \href {https://ui.adsabs.harvard.edu/abs/1997PASJ...49..367T} {49, 367}

\bibitem[\protect\citeauthoryear{{Tayal} \& {Zatsarinny}}{{Tayal} \&
  {Zatsarinny}}{2016}]{tayal:2016}
{Tayal} S.~S.,  {Zatsarinny} O.,  2016, \mn@doi [\pra]
  {10.1103/PhysRevA.94.042707}, \href
  {https://ui.adsabs.harvard.edu/abs/2016PhRvA..94d2707T} {94, 042707}

\bibitem[\protect\citeauthoryear{{Thompson}, {Nandy}, {Jamar}, {Monfils},
  {Houziaux}, {Carnochan}  \& {Wilson}}{{Thompson}
  et~al.}{1978}]{thompson:1978}
{Thompson} G.~I.,  {Nandy} K.,  {Jamar} C.,  {Monfils} A.,  {Houziaux} L.,
  {Carnochan} D.~J.,   {Wilson} R.,  1978, {Catalogue of stellar ultraviolet
  fluxes : a compilation of absolute stellar fluxes measured by the Sky Survey
  Telescope (S2/68) aboard the ESRO satellite TD-1}

\bibitem[\protect\citeauthoryear{{Tody}}{{Tody}}{1986}]{tody:1986}
{Tody} D.,  1986, in {Crawford} D.~L.,  ed.,  Society of Photo-Optical
  Instrumentation Engineers (SPIE) Conference Series Vol. 627, Instrumentation
  in astronomy VI. p.~733, \mn@doi{10.1117/12.968154}

\bibitem[\protect\citeauthoryear{{Wade}, {Donati}, {Landstreet}  \&
  {Shorlin}}{{Wade} et~al.}{2000}]{wade:2000}
{Wade} G.~A.,  {Donati} J.-F.,  {Landstreet} J.~D.,   {Shorlin} S.~L.~S.,
  2000, \mnras, \href {http://adsabs.harvard.edu/abs/2000MNRAS.313..851W} {313,
  851}

\bibitem[\protect\citeauthoryear{{Zatsarinny}}{{Zatsarinny}}{2006}]{zatsarinny:2006}
{Zatsarinny} O.,  2006, \mn@doi [Computer Physics Communications]
  {10.1016/j.cpc.2005.10.006}, \href
  {https://ui.adsabs.harvard.edu/abs/2006CoPhC.174..273Z} {174, 273}

\bibitem[\protect\citeauthoryear{{van der Bliek}, {Manfroid}  \&
  {Bouchet}}{{van der Bliek} et~al.}{1996}]{van-der-bliek:1996}
{van der Bliek} N.~S.,  {Manfroid} J.,   {Bouchet} P.,  1996, \aaps, \href
  {https://ui.adsabs.harvard.edu/abs/1996A&AS..119..547V} {119, 547}

\makeatother
\end{thebibliography}

\appendix

\newpage

\section{Observing logs, magnetic field and radial velocity measurements}

\begin{table*}
\caption{Journal of Narval spectropolarimetric observations of \her. The columns give the UT and heliocentric Julian dates of mid-exposure for each observation, the $S/N$ of Stokes $V$ spectrum per 1.8~\kms\ velocity bin at 550~nm, the rotational and orbital phases. This is followed by the Stokes $V$ signature detection flag (DD=definite detection, MD=marginal detection, ND=no detection), the mean longitudinal magnetic field, and the radial velocity measured from the hydrogen line cores. The average uncertainty of the latter measurements is 0.52~\kms. The last column contains a running number to indicate consecutive observations averaged for the purpose of magnetic mapping. Zero in this column points to the three low $S/N$ observations that were excluded from that analysis. \label{tab:obs}}
\begin{tabular}{llrcccrrr}
\hline
UT date & HJD & $S/N$ & Rotational & Orbital & Stokes $V$ & \multicolumn{1}{c}{\bz} & \multicolumn{1}{c}{$V_{\rm r}$} & $N_{\rm avg}$ \\
        &     &       & phase      & phase   & detection & \multicolumn{1}{c}{(G)} & \multicolumn{1}{c}{(\kms)} & \\
\hline
2007-03-14 & 2454173.713 & 980 & 0.000 & 0.587 & DD & $ -79\pm  7$ & $-10.4$ & 1 \\
2007-03-15 & 2454174.709 & 999 & 0.242 & 0.597 & DD & $  22\pm  7$ & $-11.2$ & 2 \\
2018-04-14 & 2458222.638 & 247 & 0.590 & 0.275 & ND & $ -47\pm 28$ & $-12.0$ & 3 \\
2018-04-15 & 2458223.518 & 433 & 0.804 & 0.284 & ND & $   4\pm 18$ & $-12.1$ & 4 \\
2018-04-19 & 2458227.646 & 434 & 0.807 & 0.326 & ND & $ -15\pm 18$ & $-11.6$ & 5 \\
2018-04-24 & 2458232.579 & 381 & 0.005 & 0.375 & ND & $ -63\pm 20$ & $-11.5$ & 6 \\
2018-04-25 & 2458233.571 & 318 & 0.246 & 0.385 & DD & $ -20\pm 24$ & $-12.2$ & 7 \\
2018-04-27 & 2458235.540 & 382 & 0.724 & 0.405 & DD & $  22\pm 20$ & $-10.4$ & 8 \\
2018-05-06 & 2458244.537 & 539 & 0.910 & 0.495 & ND & $ -58\pm 14$ & $-10.7$ & 9 \\
2018-05-09 & 2458247.513 & 477 & 0.633 & 0.525 & DD & $  61\pm 15$ & $ -9.4$ & 10 \\
2018-05-11 & 2458249.539 & 490 & 0.125 & 0.546 & ND & $ -22\pm 15$ & $-10.5$ & 11 \\
2018-05-17 & 2458255.569 & 659 & 0.590 & 0.606 & DD & $  54\pm 10$ & $ -9.4$ & 12 \\
2018-05-18 & 2458256.538 & 552 & 0.825 & 0.616 & DD & $ -56\pm 13$ & $-10.1$ & 13 \\
2018-05-18 & 2458257.487 & 655 & 0.056 & 0.626 & DD & $ -50\pm 11$ & $-10.3$ & 14 \\
2018-05-23 & 2458261.502 & 551 & 0.031 & 0.666 & ND & $ -62\pm 13$ & $-10.4$ & 15 \\
2018-06-15 & 2458284.530 & 615 & 0.625 & 0.897 & DD & $  43\pm 11$ & $-13.7$ & 16 \\
2018-06-15 & 2458284.548 & 579 & 0.630 & 0.897 & DD & $  33\pm 12$ & $-13.8$ & 16 \\
2018-06-16 & 2458285.570 & 654 & 0.878 & 0.908 & MD & $ -45\pm 11$ & $-15.3$ & 17 \\
2018-06-16 & 2458285.587 & 582 & 0.882 & 0.908 & ND & $ -25\pm 13$ & $-15.5$ & 17 \\
2018-06-17 & 2458286.556 & 523 & 0.117 & 0.918 & DD & $ -61\pm 14$ & $-16.0$ & 18 \\
2018-06-17 & 2458286.573 & 529 & 0.122 & 0.918 & DD & $ -51\pm 14$ & $-16.0$ & 18 \\
2018-06-18 & 2458287.522 & 597 & 0.352 & 0.927 & DD & $  55\pm 11$ & $-16.8$ & 19 \\
2018-06-18 & 2458287.540 & 601 & 0.356 & 0.928 & DD & $  64\pm 11$ & $-16.9$ & 19 \\
2018-06-18 & 2458288.475 & 353 & 0.583 & 0.937 & DD & $  13\pm 20$ & $-15.9$ & 20 \\
2018-06-18 & 2458288.493 & 365 & 0.588 & 0.937 & ND & $  71\pm 19$ & $-15.9$ & 20 \\
2018-06-19 & 2458289.413 & 578 & 0.811 & 0.946 & DD & $ -36\pm 13$ & $-17.2$ & 21 \\
2018-06-19 & 2458289.431 & 563 & 0.816 & 0.947 & DD & $ -49\pm 14$ & $-17.1$ & 21 \\
2018-06-24 & 2458293.511 & 541 & 0.807 & 0.988 & DD & $ -22\pm 14$ & $-18.9$ & 22 \\
2018-06-24 & 2458293.528 & 552 & 0.811 & 0.988 & DD & $ -15\pm 14$ & $-19.2$ & 22 \\
2018-06-24 & 2458294.490 & 517 & 0.045 & 0.997 & MD & $ -72\pm 14$ & $-19.6$ & 23 \\
2018-06-25 & 2458294.507 & 587 & 0.049 & 0.998 & MD & $ -52\pm 12$ & $-19.5$ & 23 \\
2018-06-25 & 2458295.461 & 586 & 0.281 & 0.007 & DD & $  37\pm 12$ & $-20.7$ & 24 \\
2018-06-25 & 2458295.480 & 488 & 0.285 & 0.007 & DD & $  52\pm 15$ & $-20.6$ & 24 \\
2018-06-30 & 2458300.459 &  72 & 0.495 & 0.057 & ND & $ 156\pm 95$ & $-18.9$ & 0 \\
2018-06-30 & 2458300.477 & 277 & 0.499 & 0.058 & ND & $  24\pm 24$ & $-18.8$ & 25 \\
2018-07-09 & 2458309.465 & 647 & 0.683 & 0.148 & DD & $  38\pm 11$ & $-15.0$ & 26 \\
2018-07-17 & 2458317.487 & 641 & 0.631 & 0.228 & DD & $  39\pm 11$ & $-12.6$ & 27 \\
2018-07-22 & 2458322.489 & 709 & 0.846 & 0.279 & MD & $ -34\pm 11$ & $-12.5$ & 28 \\
2018-07-23 & 2458323.392 & 372 & 0.066 & 0.288 & ND & $ -60\pm 20$ & $-12.4$ & 29 \\
2018-07-23 & 2458323.489 & 503 & 0.089 & 0.289 & DD & $ -88\pm 15$ & $-12.4$ & 29 \\
2018-07-25 & 2458325.446 & 394 & 0.565 & 0.308 & DD & $ 103\pm 17$ & $-11.4$ & 30 \\
2018-07-26 & 2458325.527 & 388 & 0.584 & 0.309 & DD & $  38\pm 18$ & $-11.4$ & 30 \\
2018-09-19 & 2458381.306 &  84 & 0.135 & 0.870 & ND & $-136\pm103$ & $-14.1$ & 0 \\
2018-09-23 & 2458385.297 & 473 & 0.104 & 0.910 & ND & $ -21\pm 16$ & $-15.5$ & 31 \\
2018-10-08 & 2458400.270 & 418 & 0.742 & 0.060 & DD & $ -28\pm 18$ & $-18.5$ & 32 \\
2018-10-12 & 2458404.263 & 106 & 0.711 & 0.101 & ND & $  -5\pm 77$ & $-16.1$ & 0 \\
\hline
\end{tabular}
\end{table*}

\begin{table*}
\caption{Journal of CES spectroscopic observations of \her. The columns give the UT and mid-exposure heliocentric Julian dates for each observation, exposure time, the $S/N$ at 550~nm, the rotational and orbital phases. \label{tab:obsrtt}}
\begin{tabular}{llcrcc}
\hline
UT date & HJD & Exposure & $S/N$ & Rotational & Orbital  \\
        &     & time (s) &       & phase      & phase    \\
\hline
2014-05-06 & 2456783.569 & 3000 & 168 & 0.002 & 0.814 \\
2014-05-07 & 2456785.450 & 4200 & 115 & 0.459 & 0.833 \\
2014-05-07 & 2456785.500 & 4200 & 109 & 0.472 & 0.833 \\
2014-05-11 & 2456789.436 & 4000 & 190 & 0.428 & 0.873 \\
2014-05-11 & 2456789.486 & 4000 & 206 & 0.440 & 0.873 \\
2014-05-12 & 2456789.535 & 4000 & 161 & 0.452 & 0.874 \\
2014-05-12 & 2456789.580 & 3500 & 201 & 0.463 & 0.874 \\
2014-07-10 & 2456849.277 & 4000 & 188 & 0.965 & 0.474 \\
2014-07-10 & 2456849.325 & 4000 & 241 & 0.976 & 0.475 \\
2014-07-10 & 2456849.377 & 4000 & 242 & 0.989 & 0.475 \\
2014-07-10 & 2456849.425 & 4000 & 188 & 0.001 & 0.476 \\
2014-07-10 & 2456849.469 & 3000 & 136 & 0.011 & 0.476 \\
2014-07-12 & 2456851.287 & 4000 & 198 & 0.453 & 0.494 \\
2014-07-12 & 2456851.336 & 4000 & 212 & 0.465 & 0.495 \\
2014-07-12 & 2456851.384 & 4000 & 228 & 0.476 & 0.495 \\
2014-07-12 & 2456851.432 & 4000 & 204 & 0.488 & 0.496 \\
2014-07-12 & 2456851.480 & 4000 & 224 & 0.500 & 0.496 \\
2014-08-09 & 2456879.359 & 4000 & 222 & 0.272 & 0.776 \\
2014-08-09 & 2456879.408 & 4000 & 164 & 0.284 & 0.777 \\
2014-08-10 & 2456880.264 & 4000 & 266 & 0.492 & 0.785 \\
2014-08-10 & 2456880.314 & 4000 & 218 & 0.504 & 0.786 \\
2014-08-10 & 2456880.362 & 4000 & 207 & 0.516 & 0.786 \\
2014-08-10 & 2456880.410 & 3800 & 165 & 0.528 & 0.787 \\
2014-08-11 & 2456881.261 & 4000 & 213 & 0.734 & 0.796 \\
2014-08-11 & 2456881.309 & 4000 & 196 & 0.746 & 0.796 \\
2014-08-11 & 2456881.358 & 4000 & 202 & 0.758 & 0.796 \\
2014-08-11 & 2456881.405 & 3600 & 129 & 0.769 & 0.797 \\
2015-04-28 & 2457140.542 & 1200 & 134 & 0.721 & 0.401 \\
2015-04-28 & 2457140.558 & 1200 & 130 & 0.724 & 0.401 \\
2015-04-28 & 2457140.582 & 1200 & 134 & 0.730 & 0.401 \\
2015-04-28 & 2457140.597 & 1200 & 110 & 0.734 & 0.402 \\
2015-04-28 & 2457140.613 & 1200 & 116 & 0.738 & 0.402 \\
2015-06-11 & 2457185.443 & 1800 & 113 & 0.628 & 0.852 \\
2015-06-11 & 2457185.478 & 1800 & 132 & 0.637 & 0.853 \\
2015-06-11 & 2457185.501 & 1800 & 143 & 0.642 & 0.853 \\
2015-06-12 & 2457185.523 & 1800 & 141 & 0.648 & 0.853 \\
2015-06-12 & 2457185.546 & 1800 & 122 & 0.653 & 0.853 \\
2015-08-01 & 2457236.310 & 2700 & 108 & 0.985 & 0.363 \\
2015-08-01 & 2457236.381 & 2700 & 101 & 0.002 & 0.364 \\
2015-08-01 & 2457236.416 & 2700 & 82  & 0.011 & 0.365 \\
2015-08-02 & 2457237.368 & 2700 & 120 & 0.242 & 0.374 \\
2015-08-02 & 2457237.401 & 2700 & 107 & 0.250 & 0.374 \\
2018-05-01 & 2458240.472 & 1800 & 112 & 0.922 & 0.455 \\
2018-05-01 & 2458240.496 & 1800 & 111 & 0.928 & 0.455 \\
2018-05-02 & 2458240.520 & 1800 & 114 & 0.934 & 0.455 \\
2018-05-02 & 2458240.543 & 1800 & 124 & 0.940 & 0.455 \\
2018-05-02 & 2458240.566 & 1800 & 83  & 0.945 & 0.455 \\
2018-05-02 & 2458240.590 & 1800 & 109 & 0.951 & 0.456 \\
\hline
\end{tabular}
\end{table*}

\begin{table}
\caption{Nightly mean radial velocities of \her\ measured from the H$\alpha$ line in CES observations. The columns give the heliocentric Julian date, orbital phase and radial velocity. The average uncertainty of these measurements is 0.71~\kms. \label{tab:rttrv}}
\begin{tabular}{lcc}
\hline
HJD & Orbital & $V_{\rm r}$  \\
 & phase & (\kms) \\
\hline
2456783.570 & 0.814 & $-12.8$ \\
2456785.475 & 0.833 & $-12.8$ \\  
2456789.510 & 0.873 & $-13.9$ \\  
2456849.375 & 0.475 & $ -9.5$ \\  
2456851.360 & 0.495 & $-11.7$ \\ 
2456879.383 & 0.777 & $-11.4$ \\ 
2456880.338 & 0.786 & $-11.3$ \\ 
2456881.333 & 0.796 & $-11.3$ \\ 
2457140.578 & 0.401 & $-11.3$ \\ 
2457185.499 & 0.853 & $-12.6$ \\ 
2457236.369 & 0.364 & $-11.9$ \\ 
2457237.384 & 0.374 & $-12.1$ \\ 
2458240.525 & 0.455 & $-12.2$ \\ 
\hline
\end{tabular}
\end{table}

\newpage

\section{DI line profile fits}

\begin{figure}
\centering
\figps{0.98\hsize}{0}{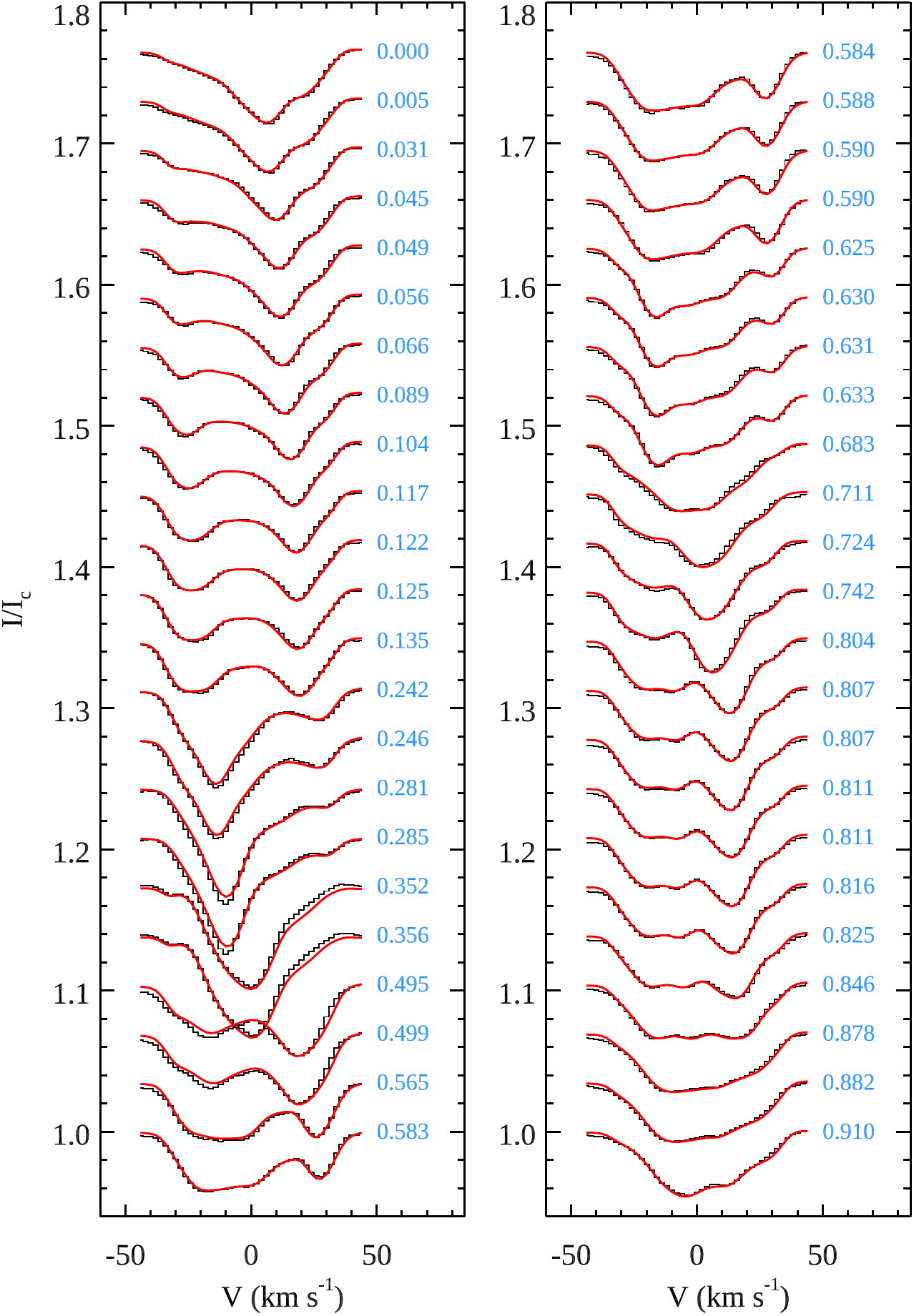}
\caption{Comparison of the observed (black histograms) and model (red solid lines) Ti Narval Stokes $I$ LSD profiles.}
\label{fig:prfti}
\end{figure}

\begin{figure}
\centering
\figps{0.98\hsize}{0}{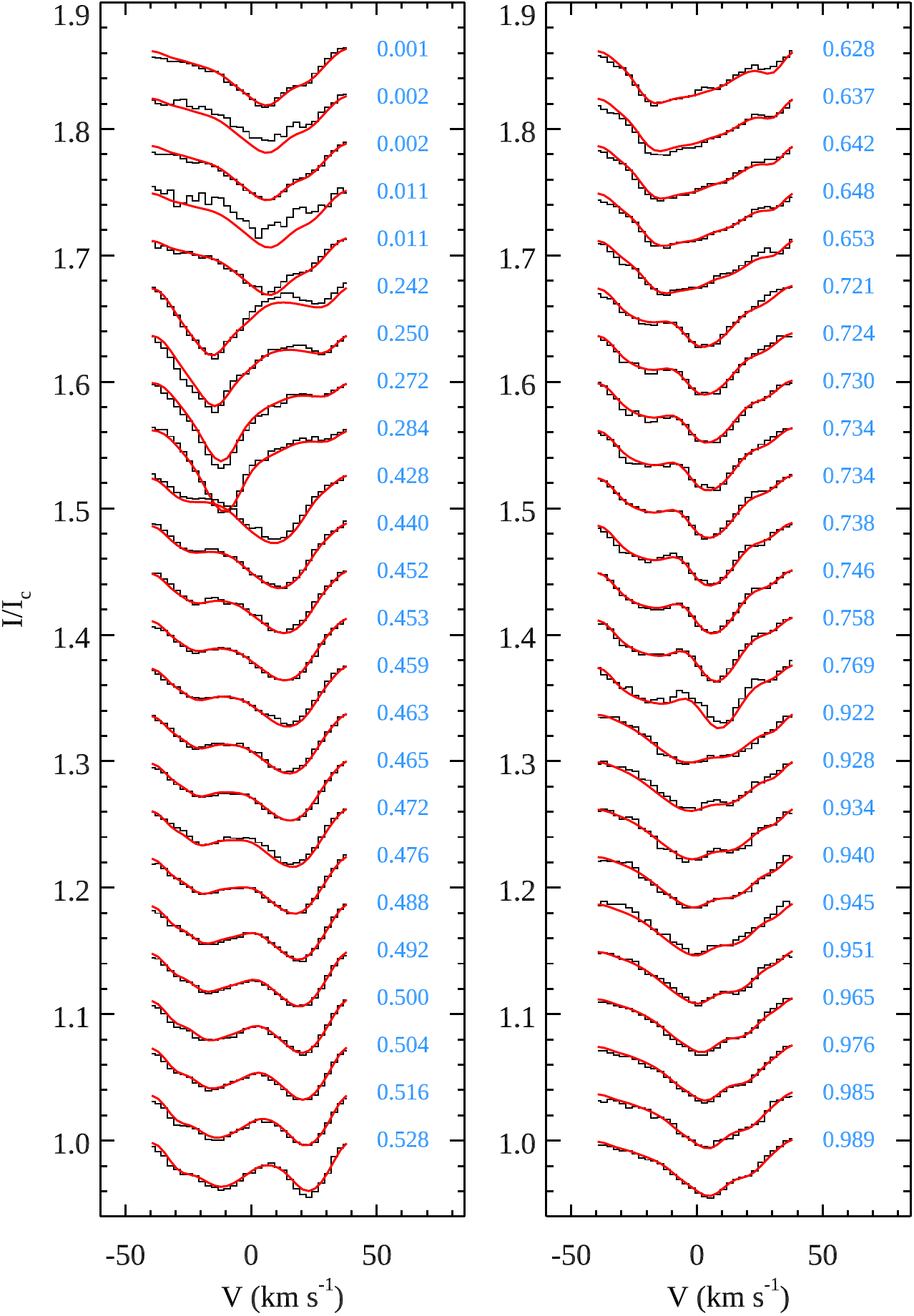}
\caption{Same as Fig.~\ref{fig:prfti} but for the Ti Stokes $I$ LSD profiles derived from CES observations.}
\label{fig:prftiT}
\end{figure}

\begin{figure}
\centering
\figps{0.98\hsize}{0}{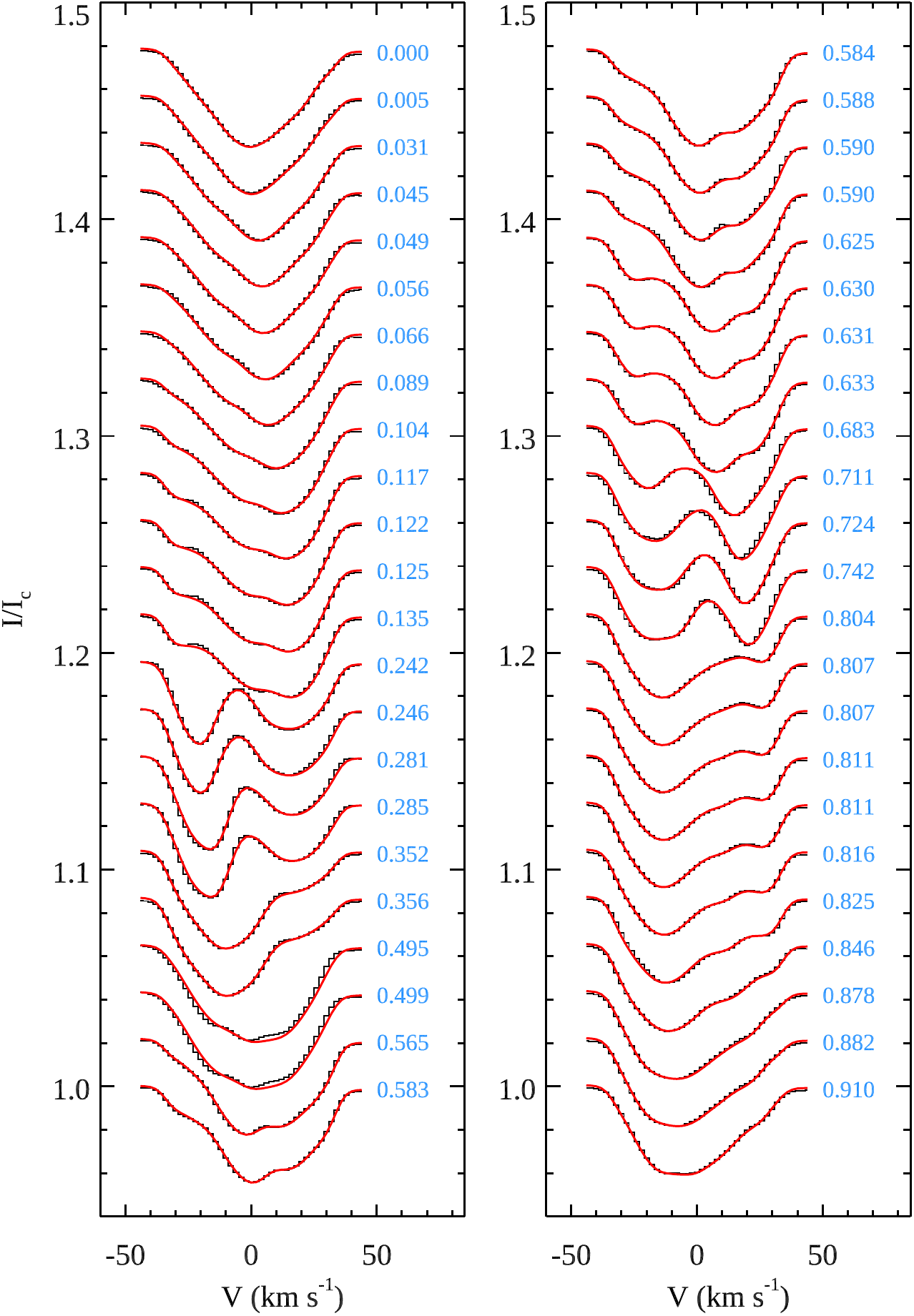}
\caption{Same as Fig.~\ref{fig:prfti}, but for Cr Narval Stokes $I$ LSD profiles.}
\label{fig:prfcr}
\end{figure}

\begin{figure}
\centering
\figps{0.98\hsize}{0}{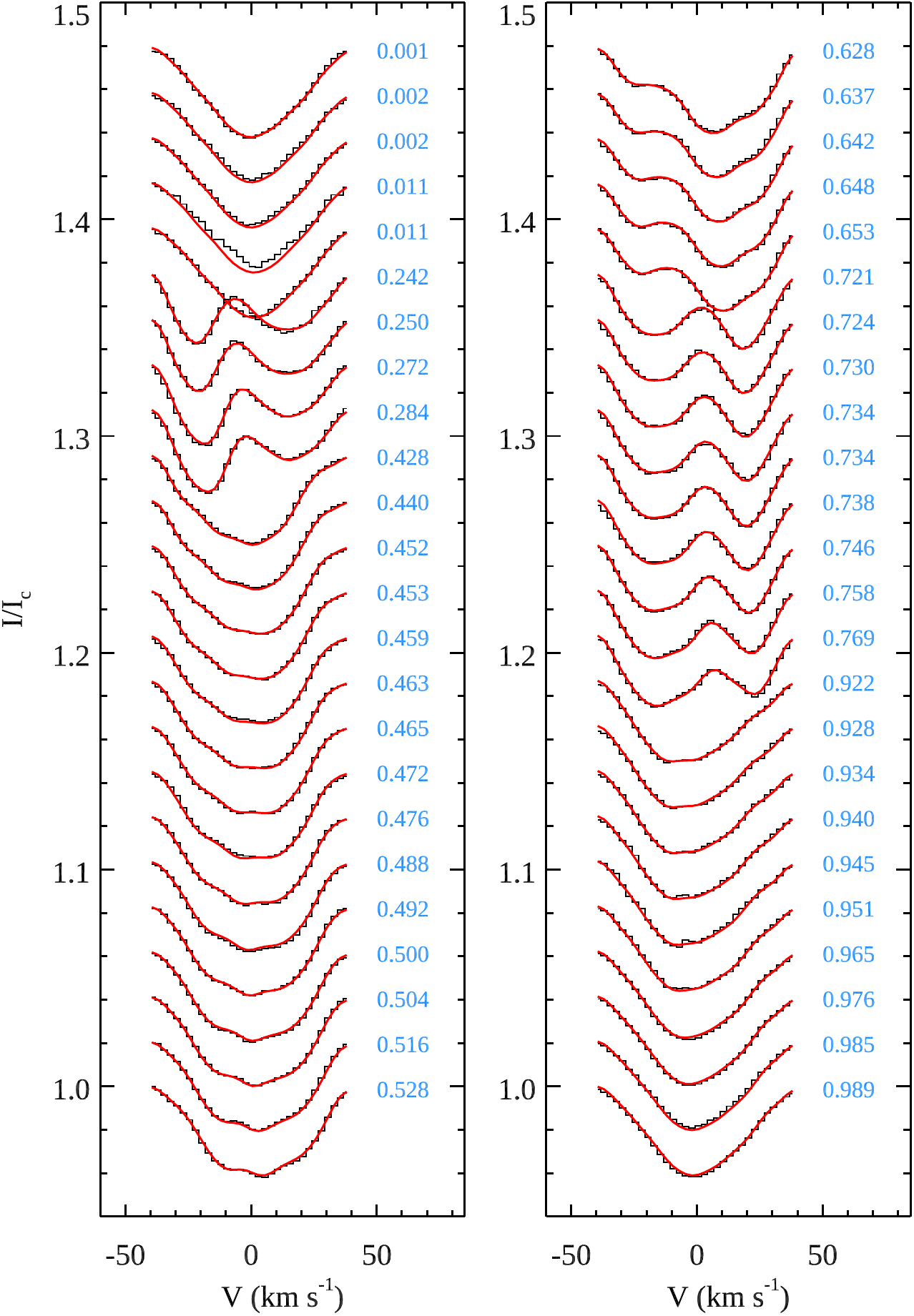}
\caption{Same as Fig.~\ref{fig:prfcr} but for the Cr Stokes $I$ LSD profiles derived from CES observations.}
\label{fig:prfcrT}
\end{figure}

\begin{figure}
\centering
\figps{0.98\hsize}{0}{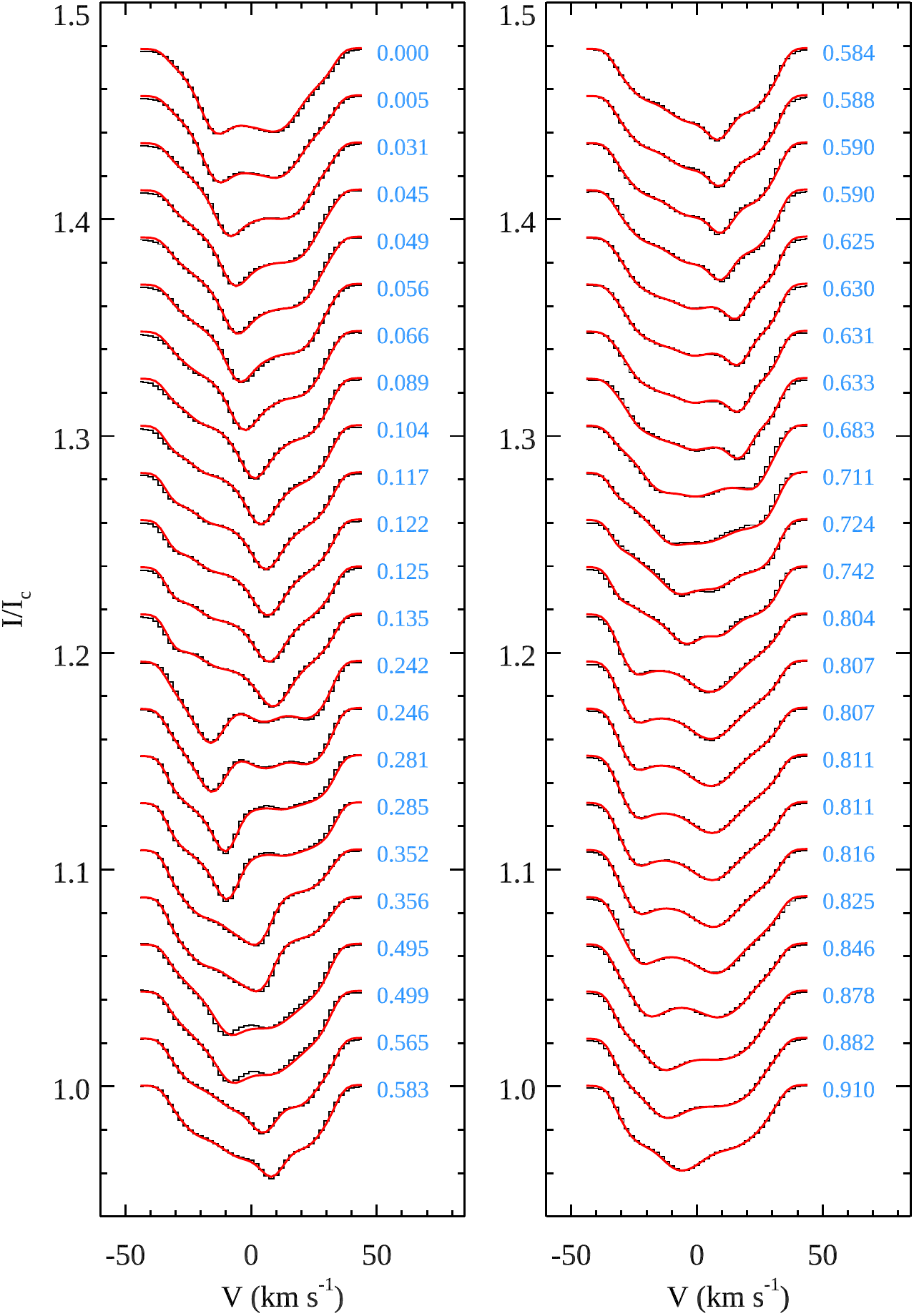}
\caption{Same as Fig.~\ref{fig:prfti}, but for Fe Narval Stokes $I$ LSD profiles.}
\label{fig:prffe}
\end{figure}

\begin{figure}
\centering
\figps{0.98\hsize}{0}{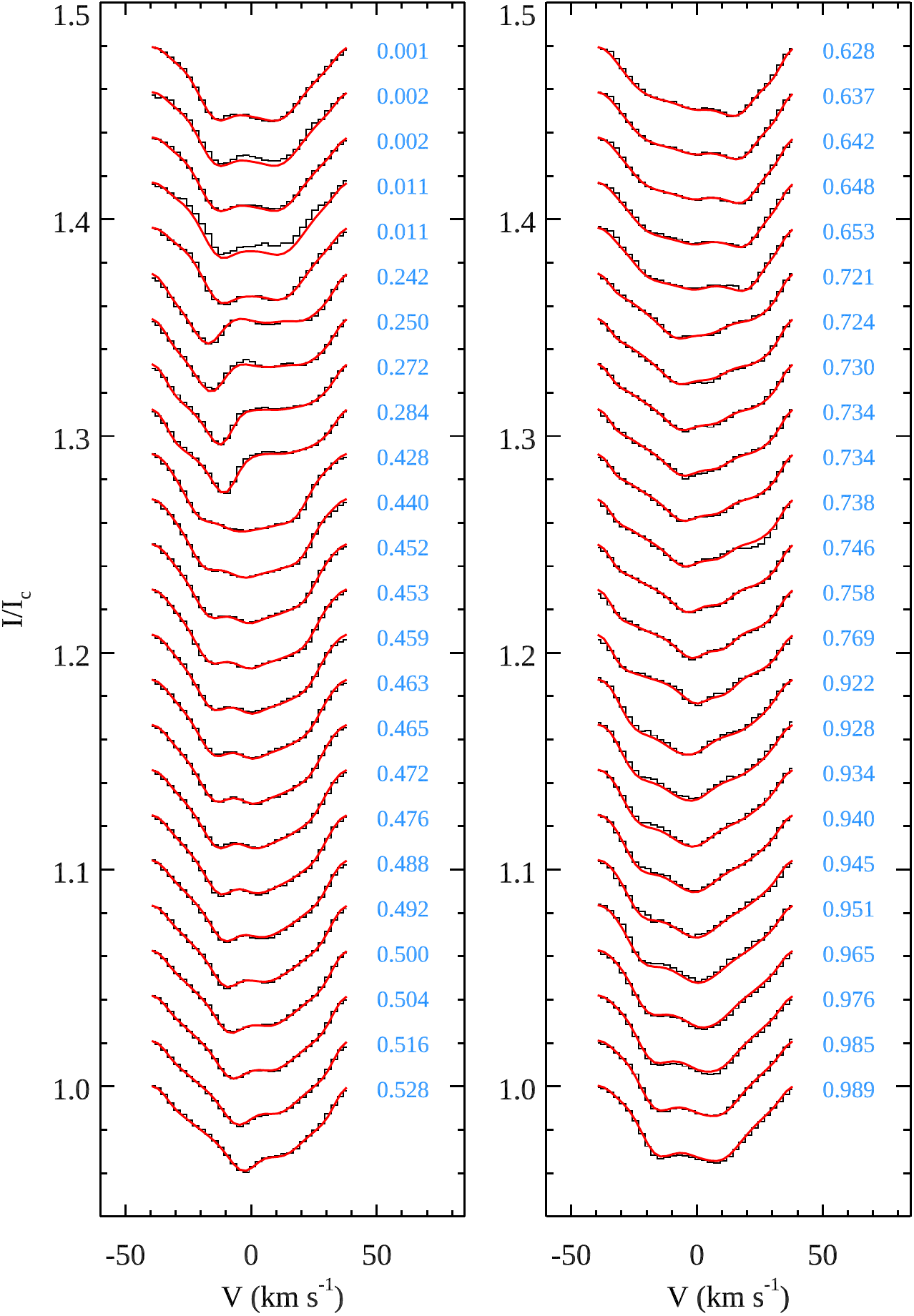}
\caption{Same as Fig.~\ref{fig:prffe} but for the Fe Stokes $I$ LSD profiles derived from CES observations.}
\label{fig:prffeT}
\end{figure}

\begin{figure}
\centering
\figps{\hsize}{0}{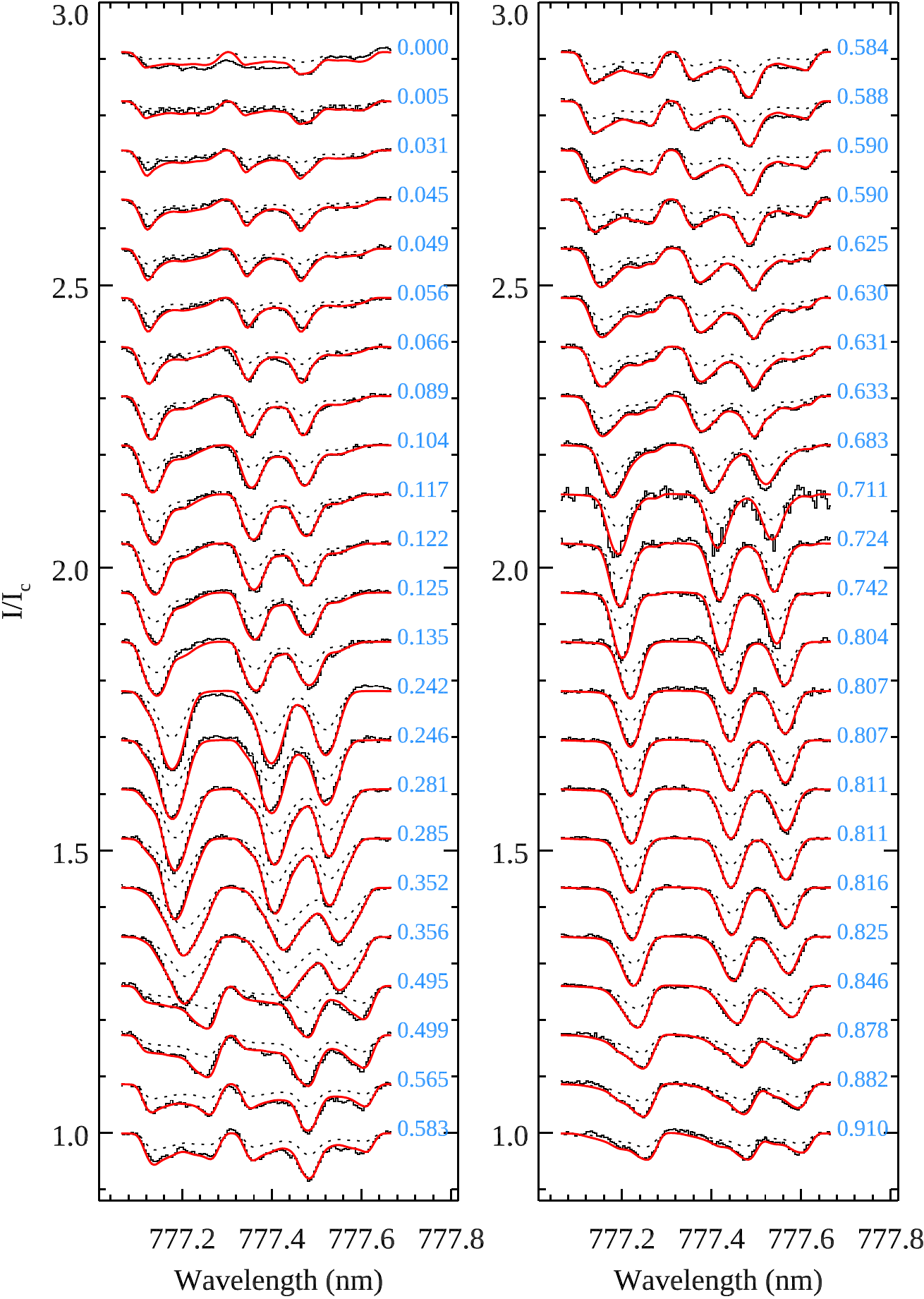}
\caption{Same as Fig.~\ref{fig:prfti}, but for the \ion{O}{i} infrared triplet in Narval spectra. Calculations accounting for NLTE effects are shown with the solid lines. The dotted lines illustrate LTE profiles for the same O surface distribution.}
\label{fig:prfo}
\end{figure}

\bsp
\label{lastpage}
\end{document}